\documentclass[ALICE,manyauthors]{cernphprep}
\usepackage[comma,square,numbers,sort&compress]{natbib}
\usepackage{amssymb}
\usepackage{amsfonts}
\usepackage{amsmath}
\usepackage{hyperref}
\usepackage{cleveref}
\usepackage{lineno}
\usepackage{xspace}
\usepackage{multirow}
\usepackage{makecell}
\usepackage{cancel}
\usepackage{booktabs}

\usepackage[T1]{fontenc}
\usepackage{orcidlink}

\newcolumntype{C}[1]{>{\centering\arraybackslash}m{#1}}

\newcommand{\dd}{\mathrm{d}}

\begin{document}
%

\newcommand{\pp}           {pp\xspace}
\newcommand{\ppbar}        {\mbox{$\mathrm {p\overline{p}}$}\xspace}
\newcommand{\XeXe}         {\mbox{Xe--Xe}\xspace}
\newcommand{\PbPb}         {\mbox{Pb--Pb}\xspace}
\newcommand{\pA}           {\mbox{pA}\xspace}
\newcommand{\pPb}          {\mbox{p--Pb}\xspace}
\newcommand{\AuAu}         {\mbox{Au--Au}\xspace}
\newcommand{\dAu}          {\mbox{d--Au}\xspace}

\newcommand{\s}            {\ensuremath{\sqrt{s}}\xspace}
\newcommand{\snn}          {\ensuremath{\sqrt{s_{\mathrm{NN}}}}\xspace}
\newcommand{\pt}           {\ensuremath{p_{\rm T}}\xspace}
\newcommand{\meanpt}       {$\langle p_{\mathrm{T}}\rangle$\xspace}
\newcommand{\ycms}         {\ensuremath{y_{\rm CMS}}\xspace}
\newcommand{\ylab}         {\ensuremath{y_{\rm lab}}\xspace}
\newcommand{\etarange}[1]  {\mbox{$\left | \eta \right |~<~#1$}}
\newcommand{\yrange}[1]    {\mbox{$\left | y \right |~<~#1$}}
\newcommand{\dndy}         {\ensuremath{\mathrm{d}N_\mathrm{ch}/\mathrm{d}y}\xspace}
\newcommand{\dndeta}       {\ensuremath{\mathrm{d}N_\mathrm{ch}/\mathrm{d}\eta}\xspace}
\newcommand{\avdndeta}     {\ensuremath{\langle\dndeta\rangle}\xspace}
\newcommand{\dNdy}         {\ensuremath{\mathrm{d}N_\mathrm{ch}/\mathrm{d}y}\xspace}
\newcommand{\Npart}        {\ensuremath{N_\mathrm{part}}\xspace}
\newcommand{\Ncoll}        {\ensuremath{N_\mathrm{coll}}\xspace}
\newcommand{\dEdx}         {\ensuremath{\textrm{d}E/\textrm{d}x}\xspace}
\newcommand{\RpPb}         {\ensuremath{R_{\rm pPb}}\xspace}

\newcommand{\nineH}        {$\sqrt{s}~=~0.9$~Te\kern-.1emV\xspace}
\newcommand{\seven}        {$\sqrt{s}~=~7$~Te\kern-.1emV\xspace}
\newcommand{\twoH}         {$\sqrt{s}~=~0.2$~Te\kern-.1emV\xspace}
\newcommand{\twosevensix}  {$\sqrt{s}~=~2.76$~Te\kern-.1emV\xspace}
\newcommand{\five}         {$\sqrt{s}~=~5.02$~Te\kern-.1emV\xspace}
\newcommand{\twosevensixnn}{$\sqrt{s_{\mathrm{NN}}}~=~2.76$~Te\kern-.1emV\xspace}
\newcommand{\fivenn}       {$\sqrt{s_{\mathrm{NN}}}~=~5.02$~Te\kern-.1emV\xspace}
\newcommand{\fivethreesixnn} {$\sqrt{s_{\mathrm{NN}}}~=~5.36$~Te\kern-.1emV\xspace}
\newcommand{\fivethreesixnnbf} {$\sqrt{\mathbf{s_{\mathrm{NN}}}}~=~\mathbf{5.36}$~\textbf{Te}\kern-.1em\textbf{V}\xspace}
\newcommand{\twohundredgevnn} {$\sqrt{s_{\mathrm{NN}}}~=~200$~Ge\kern-.1emV\xspace}
\newcommand{\LT}           {L{\'e}vy-Tsallis\xspace}
\newcommand{\GeVc}         {Ge\kern-.1emV/$c$\xspace}
\newcommand{\MeVc}         {Me\kern-.1emV/$c$\xspace}
\newcommand{\TeV}          {Te\kern-.1emV\xspace}
\newcommand{\GeV}          {Ge\kern-.1emV\xspace}
\newcommand{\MeV}          {Me\kern-.1emV\xspace}
\newcommand{\GeVmass}      {Ge\kern-.1emV/$c^2$\xspace}
\newcommand{\MeVmass}      {Me\kern-.1emV/$c^2$\xspace}
\newcommand{\lumi}         {\ensuremath{\mathcal{L}}\xspace}

\newcommand{\ITS}          {\rm{ITS}\xspace}
\newcommand{\TOF}          {\rm{TOF}\xspace}
\newcommand{\ZDC}          {\rm{ZDC}\xspace}
\newcommand{\ZDCs}         {\rm{ZDCs}\xspace}
\newcommand{\ZNA}          {\rm{ZNA}\xspace}
\newcommand{\ZNC}          {\rm{ZNC}\xspace}
\newcommand{\SPD}          {\rm{SPD}\xspace}
\newcommand{\SDD}          {\rm{SDD}\xspace}
\newcommand{\SSD}          {\rm{SSD}\xspace}
\newcommand{\TPC}          {\rm{TPC}\xspace}
\newcommand{\TRD}          {\rm{TRD}\xspace}
\newcommand{\VZERO}        {\rm{V0}\xspace}
\newcommand{\VZEROA}       {\rm{V0A}\xspace}
\newcommand{\VZEROC}       {\rm{V0C}\xspace}
\newcommand{\Vdecay} 	   {\ensuremath{V^{0}}\xspace}

\newcommand{\ee}           {\ensuremath{e^{+}e^{-}}} 
\newcommand{\pip}          {\ensuremath{\pi^{+}}\xspace}
\newcommand{\pim}          {\ensuremath{\pi^{-}}\xspace}
\newcommand{\kap}          {\ensuremath{\rm{K}^{+}}\xspace}
\newcommand{\kam}          {\ensuremath{\rm{K}^{-}}\xspace}
\newcommand{\pbar}         {\ensuremath{\rm\overline{p}}\xspace}
\newcommand{\kzero}        {\ensuremath{{\rm K}^{0}_{\rm{S}}}\xspace}
\newcommand{\lmb}          {\ensuremath{\Lambda}\xspace}
\newcommand{\almb}         {\ensuremath{\overline{\Lambda}}\xspace}
\newcommand{\Om}           {\ensuremath{\Omega^-}\xspace}
\newcommand{\Mo}           {\ensuremath{\overline{\Omega}^+}\xspace}
\newcommand{\X}            {\ensuremath{\Xi^-}\xspace}
\newcommand{\Ix}           {\ensuremath{\overline{\Xi}^+}\xspace}
\newcommand{\Xis}          {\ensuremath{\Xi^{\pm}}\xspace}
\newcommand{\Oms}          {\ensuremath{\Omega^{\pm}}\xspace}
\newcommand{\degree}       {\ensuremath{^{\rm o}}\xspace}

\newcommand{\jpsi}         {\ensuremath{{\rm J}/\psi}\xspace}
\newcommand{\psip}         {\ensuremath{\psi{\rm (2S)}}\xspace}

\newcommand{\rabs}         {\ensuremath{R_{\rm abs}}\xspace}
\newcommand{\pdca}         {\ensuremath{p \times {\rm DCA}}\xspace}
\newcommand{\fI}           {\ensuremath{f_{\rm I}}\xspace}
\newcommand{\fD}           {\ensuremath{f_{\rm D}}\xspace}
\newcommand{\fIgl}         {\ensuremath{f^{\gamma\gamma}_{\rm I}}\xspace}

\begin{titlepage}
\PHyear{2026}       
\PHnumber{143}      
\PHdate{11 May}  

\title{Exclusive dimuon production and coherent charmonium photoproduction at forward rapidity in ultra-peripheral \PbPb collisions at \fivethreesixnnbf}
\ShortTitle{Exclusive dimuons and coherent charmonium in \PbPb UPC}   

\Collaboration{ALICE Collaboration\thanks{See Appendix~\ref{app:collab} for the list of collaboration members}}
\ShortAuthor{ALICE Collaboration} 

\begin{abstract}
    This Paper presents rapidity-differential measurements of coherent \jpsi and \psip photoproduction, as well as rapidity- and mass-differential measurements of exclusive dimuon production, in the forward rapidity region $-4 < y < -2.5$ in ultra-peripheral \PbPb collisions at \fivethreesixnn using data recorded by the ALICE detector at the LHC in 2023, corresponding to an integrated luminosity of $\lumi = 1170 \pm 50~\mu{\rm b}^{-1}$.
    The \jpsi and \psip results reveal the significant role of nuclear shadowing effects. The square root of the ratio of the measured quarkonium cross section to the impulse approximation prediction is about 0.76 for \jpsi and 0.71 for \psip, at $y \approx -3$, corresponding to typical Bjorken-$x$ values of $10^{-2}$. The exclusive dimuon results highlight the sensitivity of such measurements to precise modeling of the photon flux, particularly at impact parameters near the nuclear radius.
\end{abstract}
\end{titlepage}

\setcounter{page}{2} 


\section{Introduction}
\label{sec:intro}

Ultra-peripheral heavy-ion collisions (UPC) occur when two heavy ions pass by each other at impact parameters greater than the sum of the radii of the incoming nuclei~\cite{Bertulani:2005ru}. These events are characterized by the dominant role of strong electromagnetic fields generated by the charged nuclei, which can be described using the equivalent photon approximation (EPA) and treated as fluxes of quasi-real photons with small virtuality, $q^2  < (\hslash c / R_A)^2$, where $R_A$ is the nuclear radius. The distinctive properties of UPCs provide a unique opportunity to study photon-photon and photonuclear interactions in a clean environment, in particular through vector meson photoproduction and exclusive dimuon production.

\begin{figure}[t]
    \centering
    \begin{minipage}{.49\textwidth}
        \centering
        \includegraphics[width=0.85\textwidth]{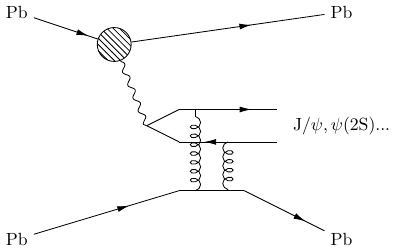}
    \end{minipage}
    \begin{minipage}{.49\textwidth}
        \centering
        \includegraphics[width=0.49\textwidth]{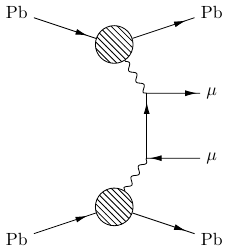}
    \end{minipage}
    \caption{Leading-order diagrams for the coherent vector meson photoproduction (left) and for the exclusive dimuon production by photon fusion (right).}
    \label{fig:feyn_diags}
\end{figure}

Vector mesons (such as \jpsi, \psip) can be produced in a photonuclear interaction either coherently or incoherently. In the former case, the incoming photon interacts with the entire nucleus, while in the latter case, it interacts with a single nucleon inside the nucleus. Coherent interaction (Fig.~\ref{fig:feyn_diags}, left) results in a lower transverse momentum (\pt) of the produced vector meson, with $\langle \pt \rangle~\sim~50$~\MeVc, while incoherent interaction leads to a higher transverse momentum with $\langle \pt \rangle~\sim~500$~\MeVc. Incoherent production is often accompanied by nuclear breakup with fragments and nucleons going in the very forward direction. In addition, an incoherent process can lead to the excitation and dissociation of the target nucleon, resulting in vector meson production at transverse momenta of $\pt \sim 1$~\GeVc and above~\cite{Guzey:2018tlk}.

In the leading-logarithmic approximation of perturbative QCD (pQCD), the coherent \jpsi and \psip photoproduction is sensitive to the gluon density in nuclei~\cite{Ryskin:1992ui,Ayala:1996em}, making these measurements valuable probes of nuclear shadowing and gluon saturation  at low Bjorken-$x$. Beyond leading order, additional contributions complicate the direct interpretation of the cross sections solely in terms of the nuclear gluon density~\cite{Flett:2019pux,Eskola:2022vpi,Flett:2024htj}.
At the Large Hadron Collider (LHC) at CERN, charmonium photoproduction measurements probe a broad Bjorken-$x$ range, $10^{-5} \lesssim x \lesssim 10^{-2}$, where $x = (m/\snn)e^{\pm y}$ for a vector meson of mass $m$ and rapidity $y$ produced in UPCs at the nucleon–nucleon center-of-mass energy \snn. These measurements therefore provide important constraints on nuclear partonic structure and enable tests of QCD-based descriptions of low-$x$ dynamics, including the shadowing and saturation models compared with the present data~\cite{Guzey:2013xba,Guzey:2016piu,Eskola:2009uj,Frankfurt:2011cs,Mantysaari:2023xcu}.

Photoproduction of \jpsi off protons has been extensively studied at HERA in electron-proton collisions~\cite{Newman:2013ada}. At the LHC, numerous results have been obtained for exclusive production of \jpsi and \psip in \pp, \pPb, and \PbPb collisions at different energies by ALICE~\cite{ALICE:2014eof,ALICE:2018oyo,ALICE:2019tqa,ALICE:2021tyx,ALICE:2021gpt,ALICE:2022wpn,ALICE:2023mfc,ALICE:2023jgu,ALICE:2023gcs,ALICE:2025cuw}, LHCb~\cite{LHCb:2013nqs,LHCb:2014acg,LHCb:2018rcm,LHCb:2022ahs,LHCb:2024pcz}, CMS~\cite{CMS:2023snh} and ATLAS~\cite{ATLAS:2025aav}. At the Relativistic Heavy Ion Collider (RHIC) at BNL, measurements of exclusive \jpsi and \psip production were performed with \AuAu UPCs at \twohundredgevnn by the STAR~\cite{STAR:2023vvb,STAR:2023nos} and PHENIX~\cite{PHENIX:2009xtn} collaborations. The exclusive \jpsi photoproduction cross section off protons measured in \pp and \pPb collisions exhibits a power-law growth with energy, showing no clear evidence of gluon saturation effects. At the same time, the suppression of coherent \jpsi photoproduction cross sections measured with respect to the impulse approximation in \PbPb UPCs can be well described by either calculations with nuclear parton distribution functions (nPDFs) or saturation-based models over a wide range of Bjorken-$x$ values, from $\sim10^{-5}$ to $\sim10^{-3}$. Precise measurements of heavy vector meson production cross sections can provide crucial constraints on theoretical models aiming to explain the observed suppression.

The production of lepton pairs is a pure QED process that proceeds via photon-photon fusion (Fig.~\ref{fig:feyn_diags}, right). The lepton-pair production cross section can be obtained by convoluting the two-photon luminosity with the leading-order elementary cross section of the Breit-Wheeler process~\cite{Breit:1934zz}. Higher-order corrections  commonly discussed in the context of dilepton production are related to the final-state radiation~\cite{Klein:2018fmp,Klein:2020jom}, Coulomb and unitarity corrections~\cite{Hencken:2006ir,Zha:2021jhf}, the latter being still a subject of theoretical debate. The two-photon luminosity can be calculated within EPA by convoluting photon fluxes emitted by the nuclei. In a widely used Monte Carlo event generator STARlight~\cite{Baltz:2009jk,Klein:2016yzr}, photon fluxes are modeled using a well-known expression for a point-like source~\cite{jackson_classical_ed}. A hard cutoff at the nuclear radius is then applied, thus neglecting the flux of photons emitted at impact parameters smaller than the nuclear radius. In a more refined approach implemented in event generators such as SuperChic 4~\cite{Harland-Lang:2018iur,Harland-Lang:2020veo} and Upcgen~\cite{Burmasov:2021phy,Burmasov:2026ytn}, photon fluxes are described using nuclear form factors 
accounting for photon impact parameters smaller than the nuclear radius, thereby allowing the possible production of lepton pairs inside nuclei.

Dilepton production measurements have been performed by ALICE in \PbPb~\cite{ALICE:2013wjo} and \pPb~\cite{ALICE:2023mfc} UPCs, by CMS~\cite{CMS:2020skx} and ATLAS~\cite{ATLAS:2020epq, ATLAS:2022srr} in \PbPb UPCs, and by PHENIX~\cite{PHENIX:2009xtn} and STAR~\cite{STAR:2004bzo,STAR:2019wlg,STAR:2023vvb} in \AuAu UPCs, with the results being generally consistent with leading-order QED calculations. 
However, recent ATLAS results~\cite{ATLAS:2020epq, ATLAS:2022srr} on dimuon and dielectron production in \PbPb UPCs show a discrepancy with STARlight calculations of up to 15--20\% at large rapidities. It has been argued that accounting for photon-nucleus impact parameters smaller than the nuclear radius may help to resolve the discrepancy~\cite{Zha:2021jhf,Burmasov:2021phy,Harland-Lang:2021ysd}. In addition, inclusion of higher-order QED contributions could improve the consistency between the data and theoretical predictions~\cite{Zha:2021jhf}.

In this Paper, we present results on coherent charmonium photoproduction and exclusive dimuon production measurements at forward rapidity $-4 < y < -2.5$ in \PbPb UPCs at a center-of-mass energy of \fivethreesixnn collected in 2023 with the upgraded ALICE detector.

\section{Experimental setup}
\label{sec:setup}

The ALICE detector~\cite{ALICE:2008ngc} at the LHC was designed and constructed with the ability to cope with the high-multiplicity environment created in collisions of heavy ions. During the LHC Long Shutdown~2~(2019--2022), the ALICE detector underwent a significant revision, enabling the full potential of the LHC. The upgrade includes updated subsystems, enhanced readout electronics, a new Central Trigger Processor, and an upgraded data acquisition system. A detailed description of the upgrades can be found in Ref.~\cite{ALICE:2023udb}. In the following, only the detector subsystems relevant for this analysis are described.

During the LHC Run 2~(2015--2018), the system of forward detectors comprising the V0 and T0 subsystems~\cite{Cortese:2004} served, among other purposes, as an interaction trigger and a luminometer, and was used to select UPCs by applying a veto on the activity associated with hadronic collisions. After the major Long Shutdown~2 upgrades, the new Fast Interaction Trigger (FIT) ~\cite{Slupecki:2022fch,Antonioli:2013ppp} has replaced the aforementioned trigger system to provide an improved performance for the high interaction rates anticipated during the LHC Run~3~(2022--2026). The FIT system includes the FV0 and FT0 detectors used for vetoing of hadronic interactions and for luminosity measurements, respectively. The FV0 detector is composed of scintillator arrays, located at $z = 3.2$~m from the interaction point on the A-side of the ALICE experiment, covering the pseudorapidity range $2.2 < \eta < 5.1$. The FT0 consists of two arrays of quartz Cherenkov radiators installed on both sides of the interaction point and covering the pseudorapidity ranges $3.5 < \eta < 4.9$ and $-3.3 < \eta < -2.1$. 

The ALICE experiment is equipped with two zero-degree calorimeters (ZDC) sensitive to the neutron emission, ZNA and ZNC~\cite{Arnaldi:1999zz,Oppedisano:2009zz}. Located at $\pm$112.5~m from the interaction point along the beam direction, they are used to detect neutral particles produced at $|\eta| > 8.8$, and also serve as luminometers for heavy-ion collisions. The calorimeters are made of a tungsten alloy with embedded quartz fibers, with a total depth of 8.7 interaction lengths. The readout system was significantly upgraded in Run 3 to sustain interaction rates of 50~kHz and higher, without dead time in continuous readout mode. The operating conditions are exceptionally challenging considering that the ZDC not only covers nucleon emission from hadronic interactions but also those resulting from electromagnetic dissociation, which have $\sim$50 times higher cross sections in \PbPb collisions at LHC energies~\cite{Pshenichnov:2001qd,Pshenichnov:2011zz,ALICE:2022xir,ALICE:2022iqi,ALICE:2024vpj}.

For the ALICE muon spectrometer, located on the C-side of the experiment at negative $z$ with respect to the nominal interaction point, significant improvements were implemented in the readout electronics~\cite{Antonioli:2013ppp} of the Muon Tracking Chambers (MCH) and the Muon Identifier~(MID)~\cite{ALICE:2008ngc}. However, the overall detector implementation has remained unchanged. The muon spectrometer provides track reconstruction and muon identification capabilities in the forward rapidity region $-4 < \eta < -2.5$. It consists of a ten interaction-length absorber followed by the five tracking stations of the MCH detector, each containing two planes of cathode pad chambers. The third station is placed inside a dipole magnet with a 3~T$\cdot$m integrated magnetic field. Downstream of the tracking stations, a 7.2 interaction-length iron wall is positioned to further reduce the remaining hadronic contamination, mainly from pion and kaon decay products. It is followed by the MID detector, constructed with four planes of resistive plate chambers.

The significant system upgrades implemented for the ALICE detector enabled the transition to the continuous readout mode and to  process collisions at high interaction rates up to 1~MHz in \pp and 50~kHz in \PbPb collisions. With the increased data sample, the precision of the coherent \jpsi cross section measurement can be improved in comparison to previous ALICE measurements at forward rapidity in \PbPb UPCs at \fivenn~\cite{ALICE:2019tqa}, enabling as well the first rapidity-differential measurement of the coherent \psip cross section. In addition, the absence of a hardware trigger on single-muon transverse momentum in the new continuous readout mode gives improved access to the exclusive dimuon production process, $\gamma\gamma\to\mu\mu$, at low invariant masses of muon pairs down to $m_{\mu\mu} \sim 1.5$~\GeVmass. However, the new data-taking regime requires a careful reconsideration of the event selection strategy applied for UPCs in the forward rapidity region.

\section{Data analysis}
\label{sec:data_analysis}

The study of coherent \jpsi and \psip photoproduction, and the exclusive dimuon production is based on a \PbPb data sample collected in 2023, corresponding to an integrated luminosity of $\lumi = 1170 \pm 50~\mu{\rm b}^{-1}$.

The integrated luminosity is estimated using reference triggers from the FT0 detector and the zero-degree calorimeters. The FT0 trigger selects events based on a signal amplitude threshold corresponding to the 50\% centrality class, where centrality is defined as the percentile of hadronic \PbPb collisions, and determined from a Glauber model fit~\cite{Loizides:2014vua,Loizides:2016djv,Loizides:2017ack}. The cross sections for these reference triggers are extrapolated from the inelastic hadronic \PbPb cross section measured at \fivenn~\cite{ALICE:2022xir} to \fivethreesixnn using the Glauber model. The ZDC triggers, used for systematic uncertainty evaluation, require the detection of neutrons on at least one side of the experiment, making them sensitive to electromagnetic dissociation and hadronic interactions of colliding nuclei. 

This analysis utilizes the muon spectrometer for track reconstruction and muon identification in the forward rapidity region. The event selection strategy requires a pair of muon tracks within the spectrometer acceptance and the absence of hadronic activity in the FV0 detector. Muon tracks are reconstructed in MCH using the tracking algorithm described in Refs.~\cite{ALICE_INT:2003_002,ALICE_INT:2009_044}. In the continuous readout mode, information from the detectors is acquired in samples corresponding to specific time intervals. The MCH readout frame (ROF) of 1~$\mu$s does not allow one to determine the timing of reconstructed MCH tracks with the required precision to resolve individual bunch crossing (BC) intervals, being about 25~ns. However, the precise timing provided by the MID detector can be used to determine the exact BC for MCH tracks matched to MID track segments (MCH-MID tracks). The MCH tracks are extrapolated to the MID detector surfaces, and the best-matching MID track segment is then selected by minimizing a $\chi^2$ function constructed from the residuals between the MCH-track and MID-segment parameters.

This analysis relies on track timing information to construct candidates, as the primary vertex position is not reconstructed for UPCs. Candidate events are formed from pairs of oppositely charged MCH-MID tracks originating from the same BC. For the studies of the MCH-MID matching uncertainty, a less strict requirement is considered: if exactly one MCH-MID track is found in a given BC, an event candidate is constructed by attaching another MCH track in a sufficiently large $\pm 10$ BC interval, covering most of the candidates.

In the constructed event candidates, each track is required to have a pseudorapidity within the muon spectrometer acceptance and a transverse momentum above $0.5$~\GeVc. In order to ensure good track quality and to reduce background contamination originating from the beam-gas interactions, the tracks are required to pass additional selection criteria. 
Limits on the track radial coordinate at the absorber end are imposed $17.6 < \rabs < 89.5$~cm in order to avoid regions of high multiple scattering. Furthermore, tracks must satisfy a selection criterion based on \pdca, which is the product of the track momentum $p$ and its distance of closest approach (DCA) to the nominal vertex at $z = 0$. This limit is set to $\pdca < 350$~cm$\times$\GeVc for tracks exiting the absorber within $17.6 < \rabs < 26.5$~cm, and $\pdca < 200$~cm$\times$\GeVc for those within $26.5 < \rabs < 89.5$~cm.

The purity of selected event candidates is ensured by applying a veto on the total FV0 signal amplitude in the bunch crossing associated with the reconstructed muon tracks. The presence of residual hadronic or electromagnetic pileup from multiple interactions per bunch crossing may result in rejection of signal events due to the FV0 veto requirement. This probability is defined as the veto inefficiency, $P_{\rm FV0}$. For the analyzed data periods with a typical interaction rate of 30 to 50~kHz, the probability of hadronic pileup did not exceed $0.35$\%. The veto inefficiency is estimated to be below $0.4$\% based on an analysis of FV0 signals from an unbiased sample of bunch crossings, selected without any additional detector requirements. The veto efficiency correction factor is estimated as a weighted average across all data-taking periods, using the corresponding integrated luminosities as weights, and is found to be $\varepsilon_{\rm veto} = 1 - P_{\rm FV0} = 99.8\%$. The associated uncertainty is discussed in Section~\ref{subsec:systematics}.

To estimate the dimuon reconstruction efficiency and model the invariant mass and transverse momentum distributions, the analysis utilizes a Monte Carlo-based full detector simulation within the ALICE O$^2$ framework~\cite{Buncic:2011297} that incorporates the continuous readout regime and a detailed description of time-dependent detector conditions for all data-taking periods. Large samples generated with STARlight 2.2~\cite{Baltz:2009jk,Klein:2016yzr} are used to simulate coherent and incoherent photoproduction of \jpsi and \psip mesons, including the feed-down process $\psip \to \jpsi + \pi\pi$. For the feed-down, \jpsi mesons are simulated assuming that they inherit the transverse polarization from the primary \psip, according to previous  measurements~\cite{BES:1999guu,ALICE:2023svb}. Exclusive dimuon production is modeled with the Upcgen generator~\cite{Burmasov:2021phy,Burmasov:2026ytn}.

\begin{figure}[tb]
    \centering
    \begin{minipage}{.49\textwidth}
        \centering
        \includegraphics[width=0.99\textwidth]{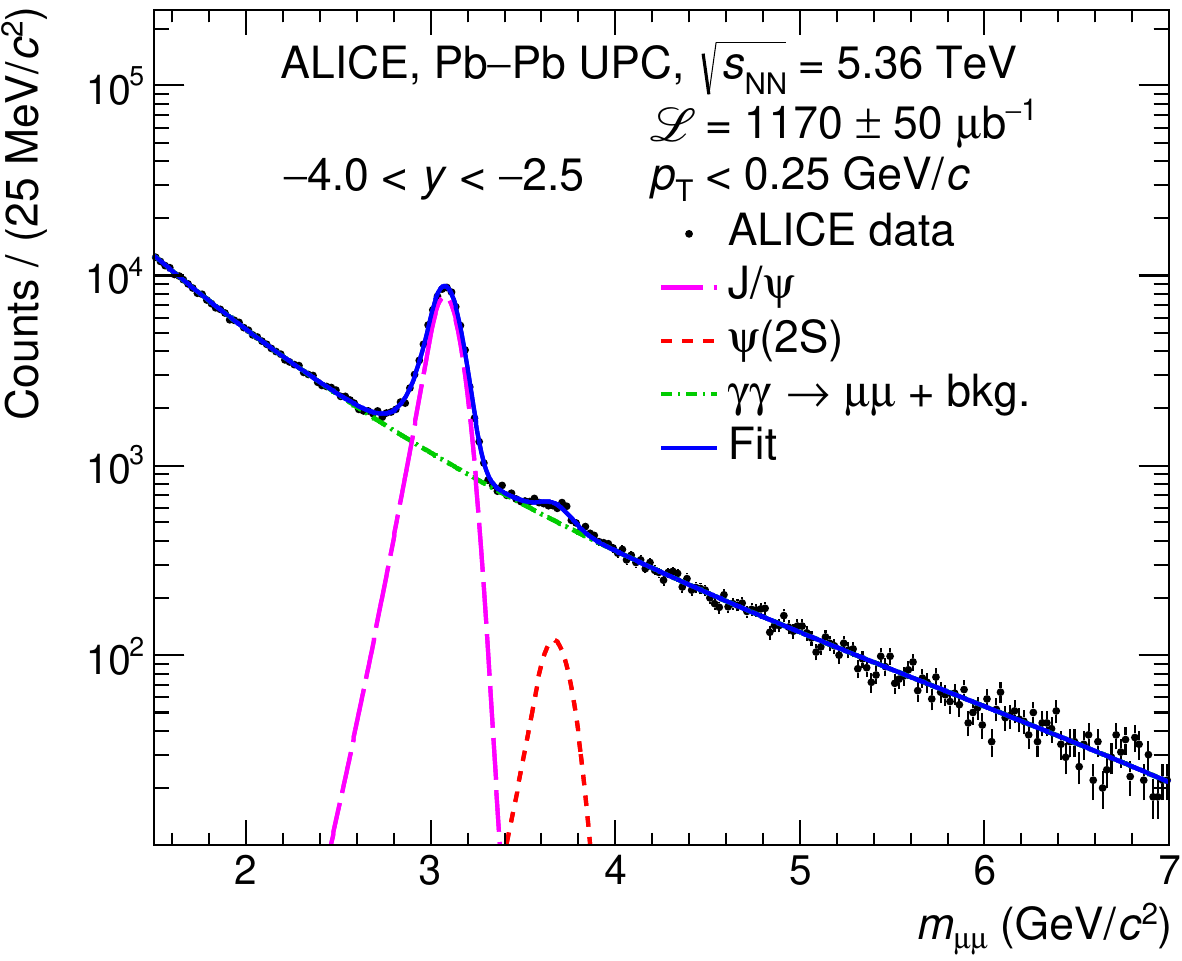}
    \end{minipage}
    \begin{minipage}{.49\textwidth}
        \centering
        \includegraphics[width=0.99\textwidth]{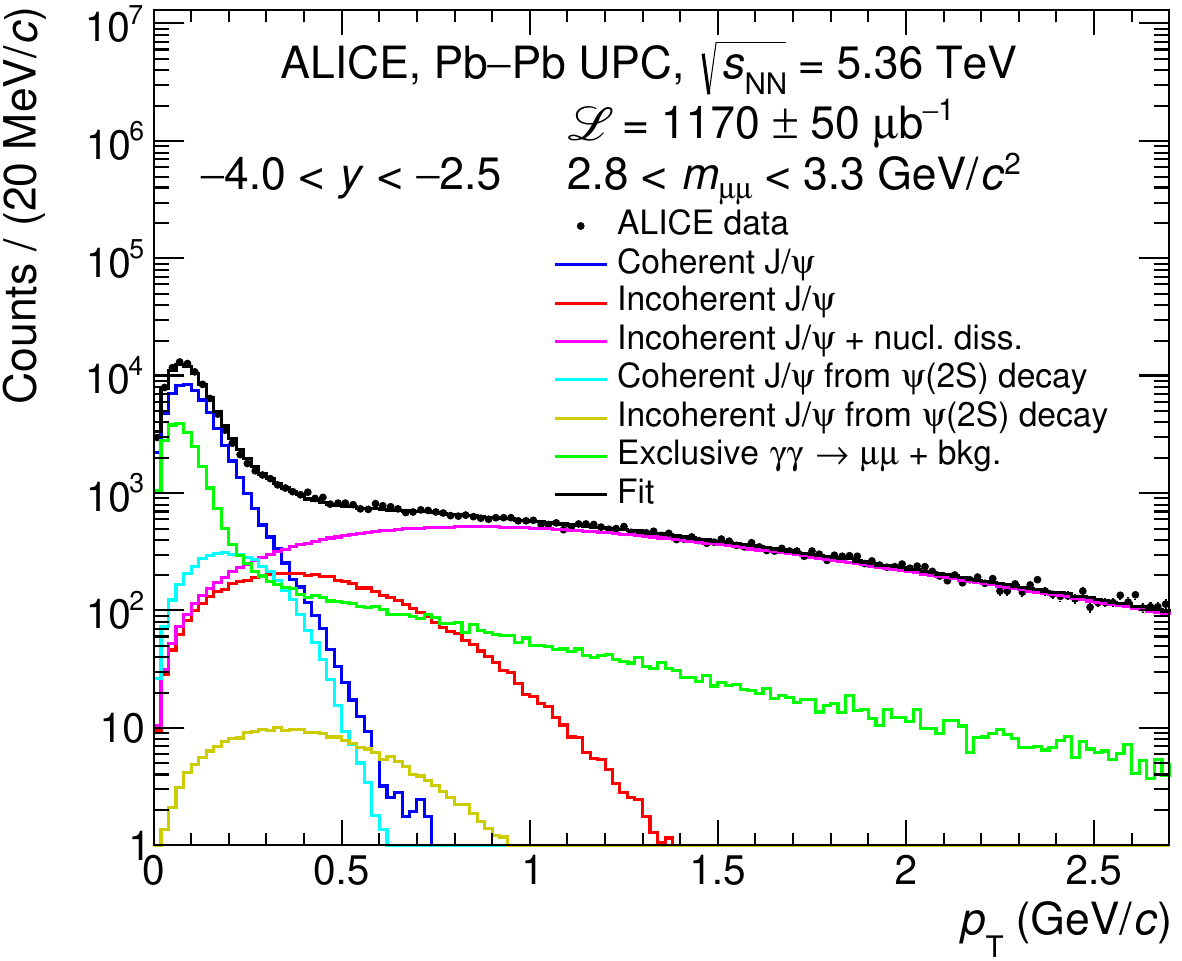}
    \end{minipage}
    \caption{Left: invariant mass distribution for the selected muon pairs with $\pt < 0.25$~\GeVc in the full forward rapidity range. Right: transverse momentum distribution for muon pairs in the invariant mass range $2.8 < m_{\mu\mu} < 3.3$~\GeVmass. Fit components are described in the text.}
    \label{fig:mass_pt_distributions_full}
\end{figure}

The \jpsi and \psip yields are measured in six and two rapidity intervals, respectively. Figure~\ref{fig:mass_pt_distributions_full},~left, shows the invariant mass distribution for the muon pairs in the full forward rapidity range after applying a selection on the dimuon transverse momentum $\pt < 0.25$~\GeVc that is imposed to suppress incoherent and background events. The fit to the mass distribution is used to extract raw inclusive \jpsi and \psip yields, as well as the raw yield for the exclusive $\gamma\gamma\to\mu\mu$ process. The shape of the non-resonant dimuon distribution is modeled with an exponential function~\cite{ALICE:2015ikg}. The peaks associated with the \jpsi and \psip mesons are fitted using double-sided Crystal Ball functions~\cite{Gaiser:1982yw,ALICE:2015ikg}. For each individual rapidity interval, the tail parameters of the Crystal Ball functions are fixed to the values obtained from the realistic simulations. The mass and width parameters of the \jpsi distributions are left unconstrained, while the \psip mass parameter is connected to the \jpsi one according to the mass difference between the two resonances obtained from the PDG~\cite{pdg:2024cfk}. Based on the ratio of their widths evaluated in simulations, the \psip width parameter is fixed to the \jpsi width multiplied by a factor of 1.1.

The fit to the invariant mass distribution shown in Fig.~\ref{fig:mass_pt_distributions_full},~left, is also used to obtain the ratio of raw inclusive \psip and \jpsi yields, $N_{\psip}$ and $N_{\jpsi}$:
\begin{equation}
    R_N = \frac{N_{\psip}}{N_{\jpsi}} = 0.017 \pm 0.002\,{\rm (stat.)} ^{+0.001}_{-0.002}\,{\rm (syst.)}\,.
    \label{eq:R_N}
\end{equation}
The associated systematic uncertainty is discussed in Section~\ref{subsec:systematics}. The measured ratio $R_N$ is then used to obtain the fraction of feed-down \jpsi in the raw inclusive \jpsi yield:
\begin{equation}
    R_{\rm fd} = \frac{N^{\rm feed-down}_{\jpsi}}{N_{\jpsi}} = \frac{(BR \times \varepsilon)_{\psip \to \jpsi + X \to \mu\mu + X}}{(BR \times \varepsilon)_{\psip \to \mu\mu}} \times R_N,
    \label{eq:R_fd}
\end{equation}
where $(BR \times \varepsilon)$ is the product of the branching ratio and reconstruction efficiency for the corresponding process. The reconstruction efficiencies were obtained using STARlight-based simulations: $\varepsilon(\jpsi \to \mu\mu) = 21\%$ and $\varepsilon(\psip \to \jpsi + X \to \mu\mu + X) = 13\%$ for the full forward rapidity range. The following branching ratios were used~\cite{pdg:2024cfk}: $BR(\psip \to \jpsi + X) = (61.5 \pm 0.7)\%$, $BR(\jpsi \to \mu\mu) = (5.961 \pm 0.033)\%$, $BR(\psip \to \mu\mu) = (0.80 \pm 0.06)\%$. 

The $R_{\rm fd}$ ratio is then used to extract the fraction of feed-down \jpsi relative to the primary \jpsi yield:
\begin{equation}
    f_{\rm D} = \frac{N^{\rm feed-down}_{\jpsi}}{N^{\rm primary}_{\jpsi}} = \frac{N^{\rm feed-down}_{\jpsi}}{N_{\jpsi} - N^{\rm feed-down}_{\jpsi}} = \left(1/R_{\rm fd} - 1\right)^{-1}.
    \label{eq:fd}
\end{equation}
The feed-down fraction $f_{\rm D} = (4.9^{+0.5}_{-0.6})\%$ is obtained for dimuon $\pt < 0.25$~\GeVc in the full forward rapidity range.
This feed-down fraction value is then used in all rapidity intervals, assuming a weak rapidity dependence, as suggested by theoretical studies of coherent \jpsi and \psip photoproduction in \PbPb UPCs~\cite{Guzey:2016piu}.

The inclusive \jpsi yields extracted from the mass fits contain contributions from both the coherent and incoherent photoproduction. In order to account for the remaining incoherent \jpsi contamination for $\pt < 0.25$~\GeVc, the \pt distribution in the dimuon invariant mass range $2.8 < m_{\mu\mu} < 3.3$~\GeVmass is fitted with templates prepared for all the processes using the simulations mentioned above, with two exceptions. For the exclusive dimuon production, side-band templates are prepared with the \pt distributions from adjacent mass regions without a charmonium signal. High-\pt contribution in the distributions, associated with incoherent \jpsi photoproduction accompanied by nucleon dissociation, is modeled with templates based on the H1 parametrization~\cite{H1:2013okq}:
\begin{equation}
    \frac{\dd N}{\dd \pt} \propto \pt \left(1 + \frac{b_{\rm pd}}{n_{\rm pd}} \pt^{2}\right)^{-n_{\rm pd}}.
    \label{eq:h1_param}
\end{equation}
In the rapidity-differential analysis, the H1 parameters are fitted in each rapidity interval individually. As an illustrative example, the fit performed over the full forward rapidity range yields $b_{\rm pd} = 0.81$~(\GeVc)$^{-2}$ and $n_{\rm pd} = 3.33$. The normalizations of the coherent and incoherent \jpsi components are left as free parameters, while the exclusive dimuon template is scaled to match the yield from the invariant mass fit. The feed-down \jpsi contribution is scaled relative to the primary \jpsi templates by the \fD fraction previously determined from the invariant mass fits in the full forward rapidity range.

The fit to the \pt distribution in the full forward rapidity range is shown in Fig.~\ref{fig:mass_pt_distributions_full},~right. The extracted contribution of the incoherent photoproduction processes, defined as $\fI = N^{\rm incoh}_{\jpsi} / N^{\rm coh}_{\jpsi}$, ranges from 3.7\% to 5.9\% for $\pt < 0.25$~\GeVc depending on the rapidity interval. The absolute value of the related systematic uncertainty varies from $-0.2$\% to $+1.5$\%, as discussed in Section~\ref{subsec:systematics}. The same \fI values are used to estimate the incoherent \psip fraction in the raw \psip yields, assuming a similar incoherent-to-coherent ratio for \psip and \jpsi~\cite{H1:2002yab,ZEUS:2022sxn}. 

\begin{figure}[tb]
    \centering
    \begin{minipage}{.49\textwidth}
        \centering
        \includegraphics[width=0.99\textwidth]{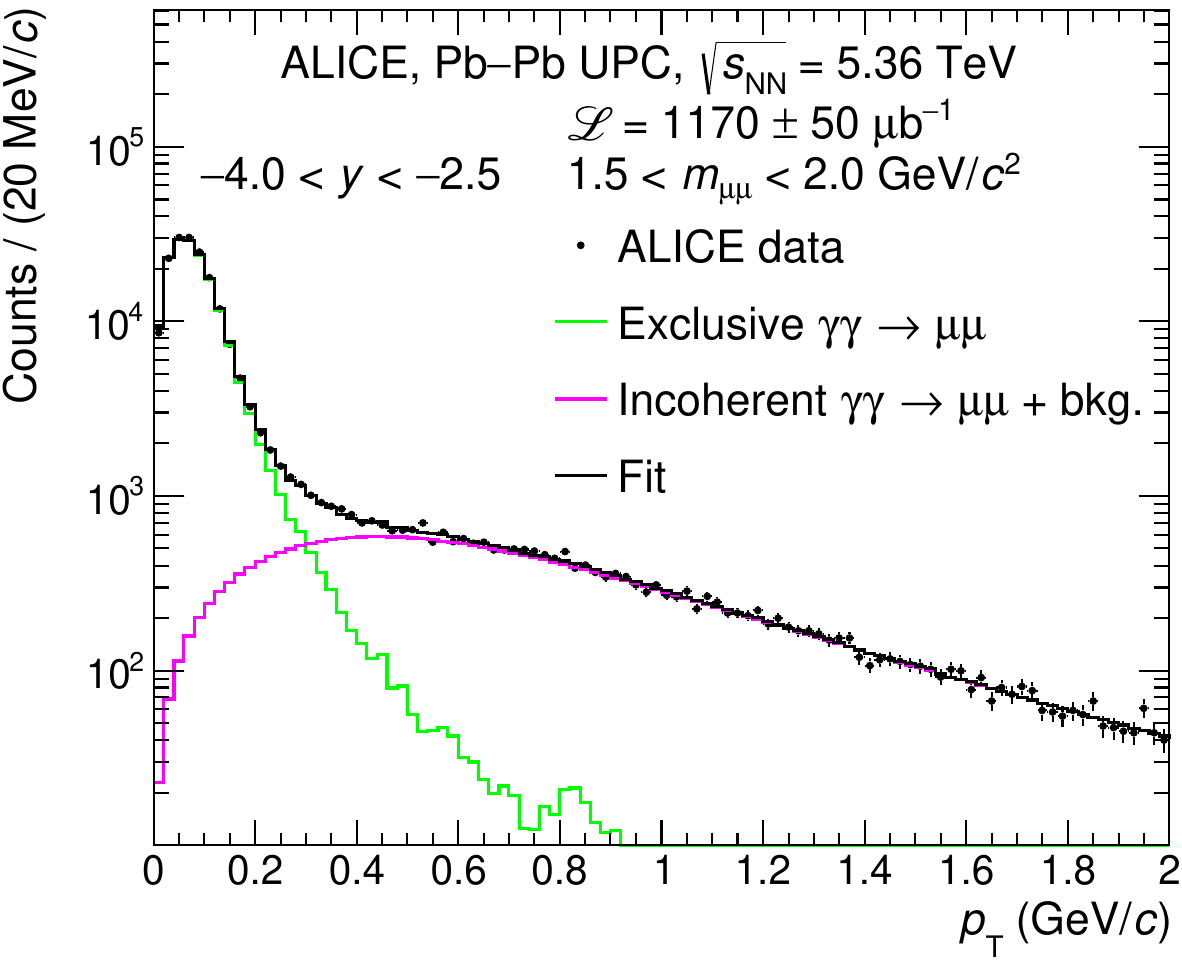}
    \end{minipage}
    \begin{minipage}{.49\textwidth}
        \centering
        \includegraphics[width=0.99\textwidth]{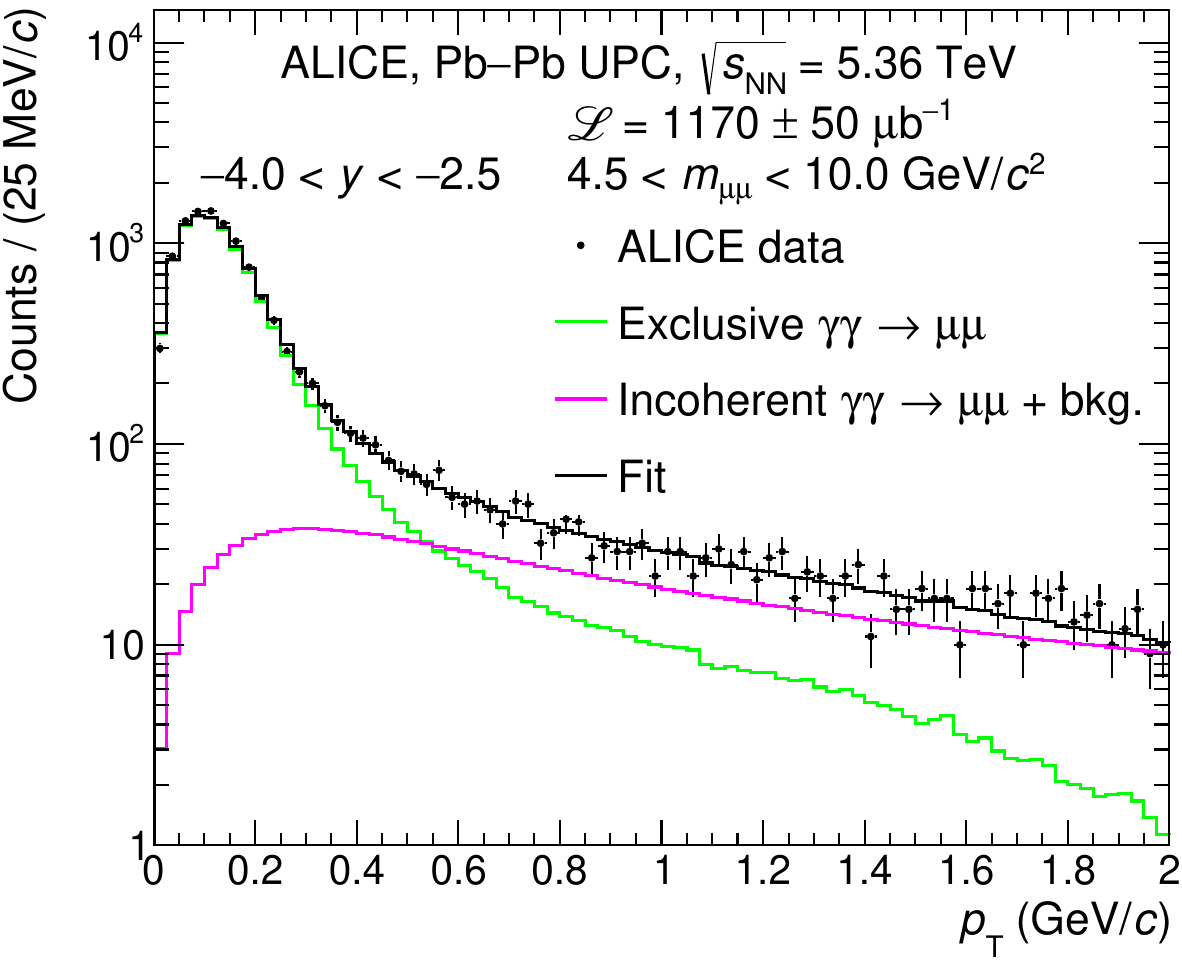}
    \end{minipage}
    \caption{Transverse momentum distribution for muon pairs in the full rapidity range for two invariant mass intervals. Fit components are described in the text.}
    \label{fig:pt_distributions_gl}
\end{figure}

Raw exclusive dimuon production yields are extracted in five invariant mass intervals, ranging from 1.5 to 10~\GeVmass, within three equal rapidity intervals. The extracted yields include several components: the exclusive dimuon production; the incoherent $\gamma\gamma\to\mu\mu$ process~\cite{Baur:1998ay}, in which at least one of the photons is emitted inelastically by the nucleus, resulting in nuclear excitation or breakup; contamination induced by beam-gas interactions and potential contamination from peripheral hadronic collisions that pass the veto requirement (discussed in Section~\ref{subsec:systematics}). The exclusive $\gamma\gamma \to \mu\mu$ signal yields, as well as the fraction of the background contributions mentioned above $\fIgl = N_{\mu\mu}^{\rm background} / N_{\mu\mu}^{\rm exclusive}$, are estimated by fitting the dimuon transverse momentum distributions in the invariant mass intervals, excluding the range $2.5 < m_{\mu\mu} < 4.5$~\GeVmass with the contribution from the \jpsi and \psip resonances. The fit utilizes the Upcgen template for the exclusive $\gamma\gamma\to\mu\mu$ and the H1 parametrization (Eq.~\ref{eq:h1_param}) for the background. To reproduce the broadening of the low-\pt component associated with the exclusive $\gamma\gamma\to\mu\mu$, a scaling factor for the width of the Upcgen template is included as a free parameter in the fit. Figure~\ref{fig:pt_distributions_gl} shows the resulting fits for the full rapidity range in the lowest and highest invariant mass intervals. The extracted background fraction ranges from 1.9\% to 6.4\% depending on the mass and rapidity interval, with the absolute systematic uncertainty varying from $-1.0$\% to $+0.6$\%. In the \jpsi and \psip invariant mass regions, a side-band method is applied where the $\fIgl$ factor is averaged over adjacent mass intervals without a charmonia signal.

\subsection{Systematic uncertainties}
\label{subsec:systematics}

A summary of systematic uncertainties assessed in this analysis is presented in Table~\ref{tab:syst_summary}.

\begin{table}[htb]
\centering
\caption{Summary of the relative systematic uncertainties, shown in percent. When two values are quoted, they correspond to the negative and positive variations obtained in the given rapidity and invariant mass intervals. The values reported in the table are the extrema of these variations observed across the analyzed intervals. Uncertainties quoted as `$\pm$' are symmetric and are taken to be identical for all intervals. Luminosity estimation, veto efficiency, and branching ratio uncertainties are fully correlated versus rapidity. All other sources of uncertainty are considered partially correlated across different rapidity and invariant mass intervals.}
\begin{tabular}{lcccccccccc}
\toprule
Source & $\sigma({\rm J /\psi})$ & $\sigma({\rm \psi(2S)})$ & $\sigma({\rm \psi(2S)}) / \sigma({\rm J /\psi})$ & $\sigma(\gamma\gamma\to\mu\mu)$ \\ \midrule \\[-0.4cm]
Luminosity                     & $\pm4.3$      & $\pm4.3$       &  --             & $\pm4.3$            \\[0.25cm]
Veto                           & $\pm1.0$      & $\pm1.0$       &  --             & $\pm1.0$            \\[0.25cm]
Tracking                       & $\pm3.0$      & $\pm3.0$       &  --             & $\pm3.0$            \\[0.25cm]
Matching                       & $+7.3$        & $+6.1$         &  --             & $+21.8$             \\[0.25cm]
Signal extraction              & $-1.1, +1.5$  & $-10.2, +5.6$  &  $-10.2, +5.8$  & $-2.3, +1.4$        \\[0.25cm]
Incoherent/background fraction & $-1.5, +0.1$  & $ -0.8, +0.1$  &  $-1.7, +0.2$   & $-0.6, +1.0$        \\[0.25cm]
Feed-down fraction             & $-0.5, +0.6$  & --             &  $-0.5, +0.6$   & --                  \\[0.25cm]
Branching ratio                & $\pm0.6$      & $\pm7.5$       &  $\pm7.5$       & --                  \\[0.25cm] \midrule
Total                          & $-5.3, +9.0$  & $-13.6, +12.3$ &  $-12.8, +9.5$  & $-5.5, +22.4$       \\
\bottomrule
\end{tabular}
\label{tab:syst_summary}
\end{table}

The 4.3\% uncertainty on the luminosity is the quadrature sum of three components: a 3.2\% uncertainty from the inelastic cross section measurement at \fivenn~\cite{ALICE:2022xir}, a 2.0\% uncertainty from the FT0 centrality determination evaluated by varying the Glauber fit anchor point, and a 2.0\% uncertainty from the variation observed when using ZDC-based triggers.

The production of vector mesons and muon pairs in peripheral heavy-ion collisions, previously observed by ALICE~\cite{ALICE:2015mzu,ALICE:2022zso}, can contaminate the UPC raw yields at the level of a few percent when the associated hadronic activity is not rejected by the FV0 veto. To estimate the magnitude of this effect, the FV0 signal threshold is varied within limits corresponding to a few percent change in the centrality of the observed collisions. The uncertainty associated with the veto efficiency is estimated to be 1\% for \jpsi, \psip, and the exclusive dimuon production.

The uncertainty of 3\% associated with the MCH tracking pseudo-efficiency stems from the accuracy of the detector description implemented in the simulation. It is extracted from comparing single-muon tracking pseudo-efficiency values obtained from simulation and data, with a procedure based on the redundancy of the tracking-chamber information, as described in Ref.~\cite{ALICE:2015jrl}. 

The systematic uncertainty on the efficiency of the MCH-MID matching originates from the evaluation of the intrinsic efficiencies of MID chambers, the algorithm used for the reconstruction of MID track segments, and the MCH-MID matching algorithm. The baseline analysis is performed using a dataset of event candidates consisting of two MCH-MID tracks. The MCH-MID matching uncertainty is evaluated using the corrected exclusive yields from event samples requiring one MCH-MID track matched to another MCH track within a $\pm 10$ BC interval. The resulting uncertainty is below $7.3\%$ for \jpsi, $6.1\%$ for \psip, and $21.8\%$ for exclusive dimuon production. The largest uncertainties are observed at low rapidities for \jpsi and \psip, and at low rapidities and low invariant masses for the exclusive dimuon production.

The uncertainties related to the \jpsi and \psip signal extraction are estimated by performing invariant mass fits with different configurations. The lower limit of the fit is varied from 1.5 to 2.5~\GeVmass, and the upper limit is varied from 5 to 7 \GeVmass. Different fitting functions for the non-resonant background are tested: the product of an exponential function and a fourth-order polynomial, an exponential function, and a variable-width Gaussian function~\cite{ALICE:2015ikg}. In addition, the \psip-to-\jpsi width ratio obtained from the simulations is varied by $\pm0.1$ based on the expected resolution in the reconstructed invariant mass of the vector mesons. To further test the invariant mass fits, the \jpsi and \psip yields are compared to the case when the tail parameters of the Crystal Ball functions are left unconstrained. The same procedure is used for the uncertainty related to the extraction of the $\gamma\gamma\to\mu\mu$ contribution in the \jpsi and \psip regions. In addition, the limits on the transverse momentum of the single muon and the value of \pdca, used for the track selection, are varied in all kinematic regions and for all the processes. The resulting signal extraction systematic uncertainties vary from $-1.1\%$ to $+1.5\%$ for \jpsi, from $-10.2\%$ to $+5.6\%$ for \psip, and from $-2.3\%$ to $+1.4\%$ for the exclusive dimuon production depending on the mass and rapidity interval. The uncertainties are asymmetric in each interval, and the quoted values represent the lowest and the highest variations observed across all intervals. For charmonia, the dominant contribution to the signal extraction uncertainty originates from the choice of the background function, while for the exclusive dimuon production, the main source of uncertainty is due to variation of the \pdca selection criterion.

In order to estimate systematic uncertainties associated with \pt template fits, several tests are performed. For the \jpsi analysis, systematic tests include varying the invariant mass range for the transverse momentum fits (with the lower limit varied between 2.5 and 2.8 \GeVmass and the upper one between 3.2 and 3.5 \GeVmass), and varying the single-muon selection criteria for transverse momentum and \pdca. The uncertainty associated with extracting the incoherent and background fractions is estimated to be between $-1.5\%$ and $+0.1\%$ for \jpsi, and between $-0.8\%$ to $+0.1\%$ for \psip. For the exclusive dimuon production, the variations result in an uncertainty ranging from $-0.6\%$ to $+1.0\%$ depending on the rapidity and invariant mass interval.

An additional uncertainty arises from the \pt fitting procedure due to the estimation of the feed-down \jpsi fraction, affecting both the \jpsi cross section and the \psip-to-\jpsi yield ratio. The feed-down fraction \fD is varied from $4.3\%$ to $5.4\%$, based on the systematic uncertainty of the \psip-to-\jpsi yield ratio (Eq.~\ref{eq:R_N}) and the associated branching ratio uncertainties (Eq.~\ref{eq:R_fd}). This variation leads to an uncertainty ranging from $-0.5\%$ to $+0.6\%$, depending on the rapidity interval.

Variations of the generated rapidity distributions in the simulations are found to have a negligible impact on the corrected yields, and therefore, this contribution is not included in the uncertainty evaluation. Possible migrations between neighboring rapidity and invariant-mass intervals due to detector resolution are also found to be negligible.

The checks relevant to the \jpsi and \psip analyses are repeated for the measurement of the \psip-to-\jpsi cross section ratio. The uncertainties arising from the \psip signal extraction are found to be dominant. The uncertainties stemming from the luminosity determination, veto efficiency estimates, evaluation of the MCH tracking and MCH-MID matching efficiencies largely cancel in the ratio measurement.

\section{Results and discussion}
\label{sec:results}

\subsection{Coherent \texorpdfstring{\jpsi}{J/Psi} and \texorpdfstring{\psip}{Psi(2S)} cross sections}
\label{subsec:results_vm}

The rapidity-differential cross section of coherent \jpsi and \psip is obtained for a given rapidity interval $\Delta y$ with
\begin{equation}
    \frac{\dd \sigma^{\rm coh}_{\rm VM}}{\dd y} = \frac{N^{\rm excl}_{\rm VM}}{\varepsilon_{\rm rec} \varepsilon_{\rm veto} BR({\rm VM} \to \mu\mu) \lumi \Delta y}~,
    \label{eq:cs_vm}
\end{equation}
where $\varepsilon_{\rm rec}$ denotes the reconstruction efficiency of the produced vector meson $\rm VM$, and $N_{\rm VM}^{\rm excl}$ is the estimated exclusive yield that can be obtained from the corresponding raw yield with $N^{\rm excl}_{\rm VM} = N_{\jpsi}/(1 + \fI + \fD)$ for the coherent \jpsi and with $N^{\rm excl}_{\rm VM} = N_{\psip}/(1 + \fI)$ for the coherent \psip, respectively. The raw \jpsi and \psip yield values, reconstruction efficiencies, \fI fractions, and coherent \jpsi and \psip cross sections with corresponding statistical and systematic uncertainties are summarized in \Cref{tab:jpsi_summary_total,tab:psi2s_summary_total}, respectively.

\begin{table}[h]
\centering
\caption{Raw ${\jpsi}$ yields with their statistical uncertainties, \jpsi reconstruction efficiencies, $f_{\rm I}$ fractions, and coherent cross sections.}
\begin{tabular}{C{2.5cm}C{2.5cm}C{1.5cm}C{1.5cm}C{5cm}} \toprule
Rapidity & $N_{\rm J/\psi}$ & $\varepsilon_{\rm rec}$~(\%) & $f_{\rm I}$ & $\dd\sigma/\dd y$~(mb) \\ \midrule
$(-4.00, -2.50)$ & $67548 \pm 346$ & $21$   & $0.052$ & $2.78 \pm 0.01$~(stat.)~$_{-0.15}^{+0.19}$~(syst.) \\ \midrule
$(-4.00, -3.75)$ & $2471  \pm 61$  & $6.8$  & $0.059$ & $1.89 \pm 0.05$~(stat.)~$_{-0.10}^{+0.11}$~(syst.) \\ \addlinespace[0.25cm]
$(-3.75, -3.50)$ & $9578  \pm 123$ & $22$   & $0.058$ & $2.28 \pm 0.03$~(stat.)~$_{-0.12}^{+0.13}$~(syst.) \\ \addlinespace[0.25cm]
$(-3.50, -3.25)$ & $18577 \pm 180$ & $36$   & $0.053$ & $2.67 \pm 0.03$~(stat.)~$_{-0.15}^{+0.16}$~(syst.) \\ \addlinespace[0.25cm]
$(-3.25, -3.00)$ & $20057 \pm 190$ & $35$   & $0.051$ & $2.97 \pm 0.03$~(stat.)~$_{-0.16}^{+0.24}$~(syst.) \\ \addlinespace[0.25cm]
$(-3.00, -2.75)$ & $13081 \pm 153$ & $22$   & $0.041$ & $3.20 \pm 0.04$~(stat.)~$_{-0.17}^{+0.26}$~(syst.) \\ \addlinespace[0.25cm]
$(-2.75, -2.50)$ & $3586  \pm 83$  & $5.6$  & $0.037$ & $3.38 \pm 0.08$~(stat.)~$_{-0.19}^{+0.31}$~(syst.) \\ \bottomrule
\end{tabular}
\label{tab:jpsi_summary_total}
\end{table}

\begin{table}[h]
\centering
\caption{Raw ${\psi({\rm 2S})}$ yields with their statistical uncertainties, \psip reconstruction efficiencies, $f_{\rm I}$ fractions and coherent cross sections.}
\begin{tabular}{C{2.5cm}C{2.5cm}C{1.5cm}C{1.5cm}C{5cm}} \toprule
Rapidity & $N_{\rm \psi(2S)}$ & $\varepsilon_{\rm rec}$~(\%) & $f_{\rm I}$ & $\dd\sigma/\dd y$~(mb) \\ \midrule
$(-4.00, -2.50)$ & $1164 \pm 105$ & $22$ & $0.052$ & $0.36 \pm 0.03$~(stat.)~$_{-0.04}^{+0.04}$~(syst.) \\ \midrule
$(-4.00, -3.25)$ & $503 \pm 66$   & $24$ & $0.056$ & $0.29 \pm 0.04$~(stat.)~$_{-0.04}^{+0.03}$~(syst.) \\ \addlinespace[0.25cm]
$(-3.25, -2.50)$ & $638 \pm 81$   & $20$ & $0.047$ & $0.43 \pm 0.05$~(stat.)~$_{-0.04}^{+0.05}$~(syst.) \\ \bottomrule
\end{tabular}
\label{tab:psi2s_summary_total}
\end{table}

\begin{figure}[htb]
    \centering
    \begin{minipage}{.49\textwidth}
        \centering
        \includegraphics[width=0.99\textwidth]{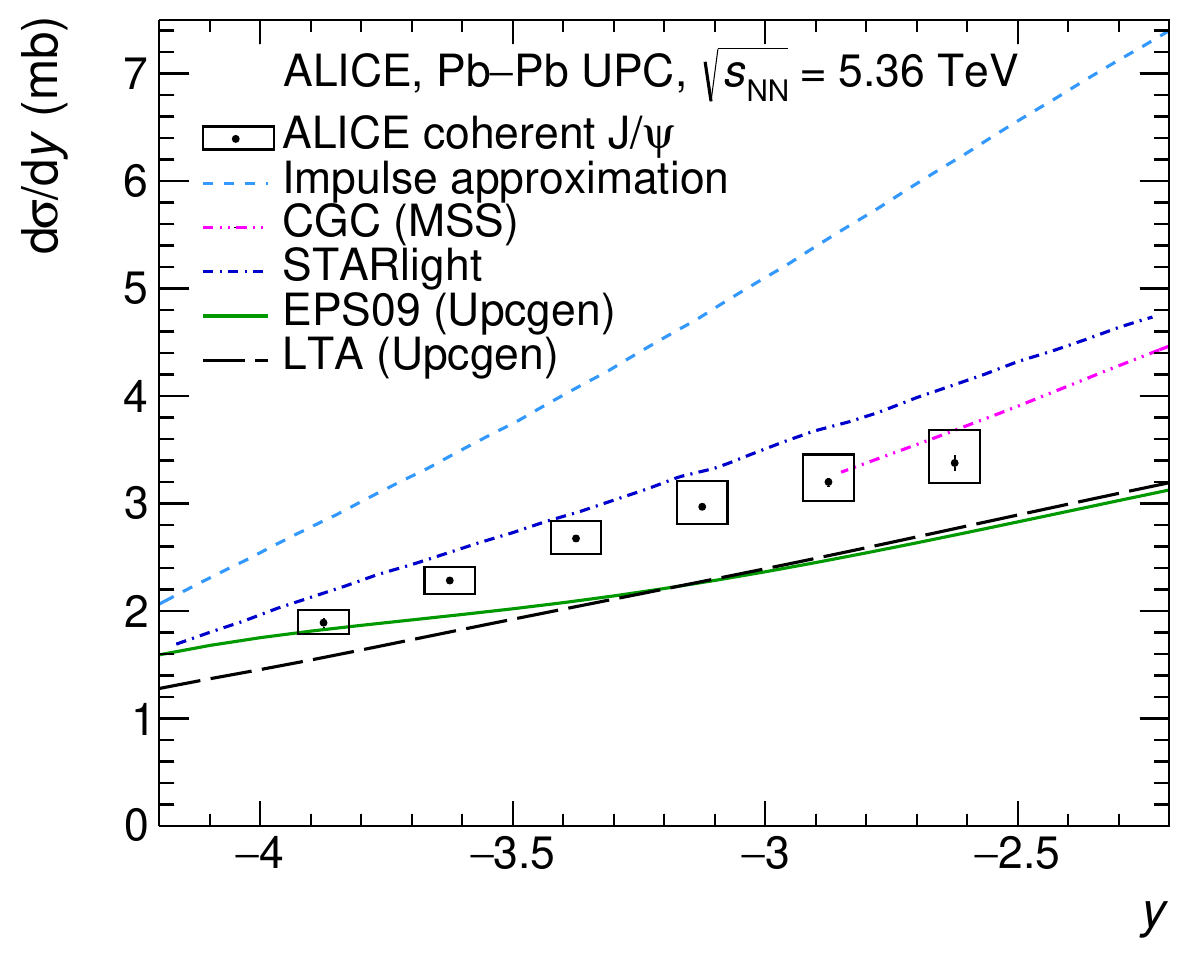}
    \end{minipage}
    \begin{minipage}{.49\textwidth}
        \centering
        \includegraphics[width=0.99\textwidth]{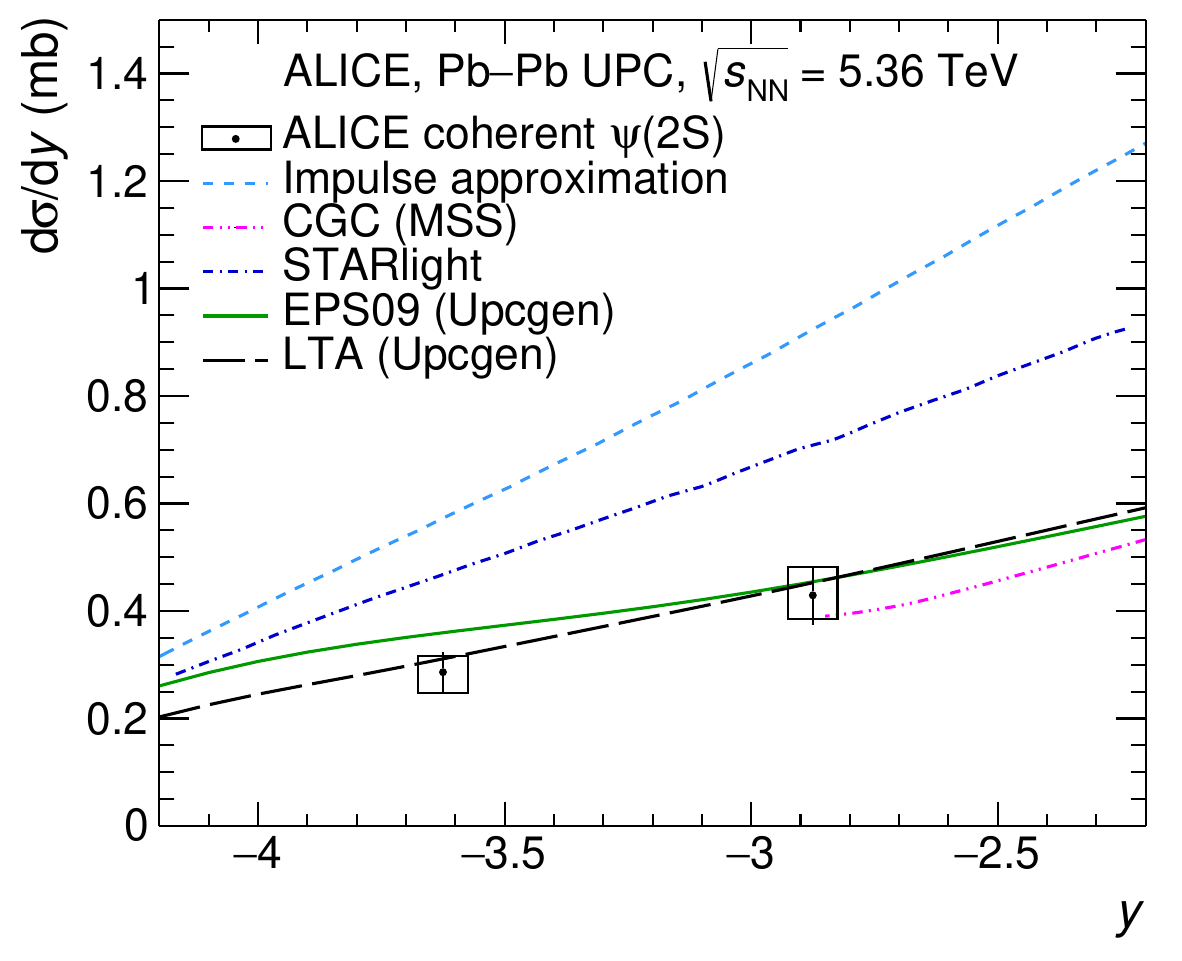}
    \end{minipage}
    \caption{Measured rapidity-differential cross sections of the coherent \jpsi~(left) and \psip~(right) production in \PbPb UPCs at \fivethreesixnn. The results are compared with theoretical calculations described in the text. The statistical uncertainties are shown with bars, the systematic uncertainties are shown with boxes.}
    \label{fig:cs_vm}
\end{figure}

The results for coherent \jpsi and \psip cross sections as a function of rapidity are presented in Fig.~\ref{fig:cs_vm} and are compared with theoretical predictions. The \jpsi measurements probe the gluon content of the nucleus at Bjorken-$x$ values either in the range $1.1 \times 10^{-5} < x < 4.7 \times 10^{-5}$ or $0.7 \times 10^{-2} < x < 3.2 \times 10^{-2}$, depending on which nucleus emitted the photon. The corresponding Bjorken-$x$ ranges for \psip are $1.3 \times 10^{-5} < x < 5.6 \times 10^{-5}$ and $0.8 \times 10^{-2} < x < 3.8 \times 10^{-2}$. 

The impulse approximation, taken from STARlight~\cite{Klein:1999qj,Baltz:2009jk,Klein:2016yzr}, is based on the HERA data for the exclusive \jpsi photoproduction off protons and neglects all nuclear effects except for coherence. The square root of the ratio of experimental points to the impulse approximation cross section is about $0.76$ for \jpsi and $0.71$ for \psip, reflecting the magnitude of the nuclear gluon shadowing factor at $y \approx -3$ corresponding to typical Bjorken-$x$ values around $10^{-2}$, under the assumption that the contribution from low Bjorken-$x$ values $\sim 10^{-5}$ can be neglected~\cite{Guzey:2013xba}.

STARlight predictions for coherent vector meson cross sections in Pb–Pb UPCs are based on the Vector Meson Dominance model for the conversion of photons into vector mesons and a Glauber-like formalism to account for multiple scattering effects. STARlight overpredicts both the \jpsi and \psip cross sections, with a larger discrepancy for the latter, suggesting that the observed suppression of the coherent vector meson cross sections cannot be fully explained by Glauber-like rescatterings alone, and other effects, such as gluon shadowing, must be considered.

Gluon shadowing effects are taken into account explicitly in the predictions from the Upcgen generator~\cite{Burmasov:2021phy,Burmasov:2026ytn} that is based on the leading-order perturbative QCD approach introduced in Ref.~\cite{Guzey:2016piu}. Two gluon shadowing models are considered: the first is based on the EPS09 LO parametrization of the available nuclear shadowing data~\cite{Eskola:2009uj} (EPS09) and the other on the leading twist approximation (LTA) of nuclear shadowing~\cite{Frankfurt:2011cs} with the weak shadowing scenario. The Upcgen predictions are found to be in good agreement with \psip cross section measurements. However, both shadowing models lie below the measured \jpsi cross section in the range $-3.75 < y < -2.5$.

Predictions from M{\"a}ntysaari, Salazar, and Schenke (MSS) are based on the color glass condensate (CGC) implementation of saturation effects accounting for event-by-event fluctuations of the nucleon substructure in heavy nuclei~\cite{Mantysaari:2023xcu}. The model is only valid for Bjorken-$x$ smaller than $10^{-2}$ corresponding to $|y| < 2.85$. Within this domain, the model predictions agree with the measured \jpsi and \psip cross sections. In addition, the model provides a reasonable description of the recent ATLAS measurements at \fivethreesixnn~\cite{ATLAS:2025aav}, although some tension remains between the ATLAS results and ALICE midrapidity measurements at \fivenn~\cite{ALICE:2021gpt}.

The obtained results can be compared with the previous measurement performed by ALICE in the same rapidity region in \PbPb UPCs at \fivenn~\cite{ALICE:2019tqa}. The reported value for the coherent \jpsi cross section measured across the full rapidity interval is $2.549 \pm 0.022{\rm (stat.)} ^{+0.209}_{-0.237}{\rm (syst.)}$ mb, which is 9\% lower than the value reported in this Paper. This difference is consistent, within the experimental uncertainties, with theoretical expectations based on STARlight's implementation of the impulse approximation, which predicts a 6--11\% increase depending on the rapidity interval.

The \psip-to-\jpsi cross section ratio in the full forward rapidity range and in two rapidity intervals is presented in Table~\ref{tab:ratio_summary_total}. The measured \psip-to-\jpsi cross section ratio agrees within 1.1$\sigma$ with previous ALICE~\cite{ALICE:2019tqa} and LHCb~\cite{LHCb:2022ahs} measurements at forward rapidity in \PbPb UPCs at \fivenn: $R = 0.150 \pm 0.018 {\rm (stat.)} \pm 0.021 {\rm (syst.)} \pm 0.007 {\rm (BR)}$ and $R = 0.155 \pm 0.014 {\rm (stat.)} \pm 0.003 {\rm (syst.)}$, respectively.

\begin{table}[thb]
\centering
\caption{Ratio of coherent ${\rm J}/\psi$ and \psip cross sections in three rapidity intervals.}
\begin{tabular}{C{2.5cm}C{6cm}}
\toprule
Rapidity & $\sigma({\rm \psi(2S)}) / \sigma({\rm J/\psi})$ \\ \midrule \\[-0.4cm]
$(-4.00, -2.50)$ & $0.131 \pm 0.012$~(stat.)~$_{-0.013}^{+0.013}$~(syst.) \\ \midrule
$(-4.00, -3.25)$ & $0.124 \pm 0.016$~(stat.)~$_{-0.016}^{+0.011}$~(syst.) \\ \addlinespace[0.25cm]
$(-3.25, -2.50)$ & $0.132 \pm 0.017$~(stat.)~$_{-0.011}^{+0.015}$~(syst.) \\ \bottomrule
\end{tabular}
\label{tab:ratio_summary_total}
\end{table}

\begin{figure}[htb]
    \centering
    \begin{minipage}{.49\textwidth}
        \centering
        \includegraphics[width=0.99\textwidth]{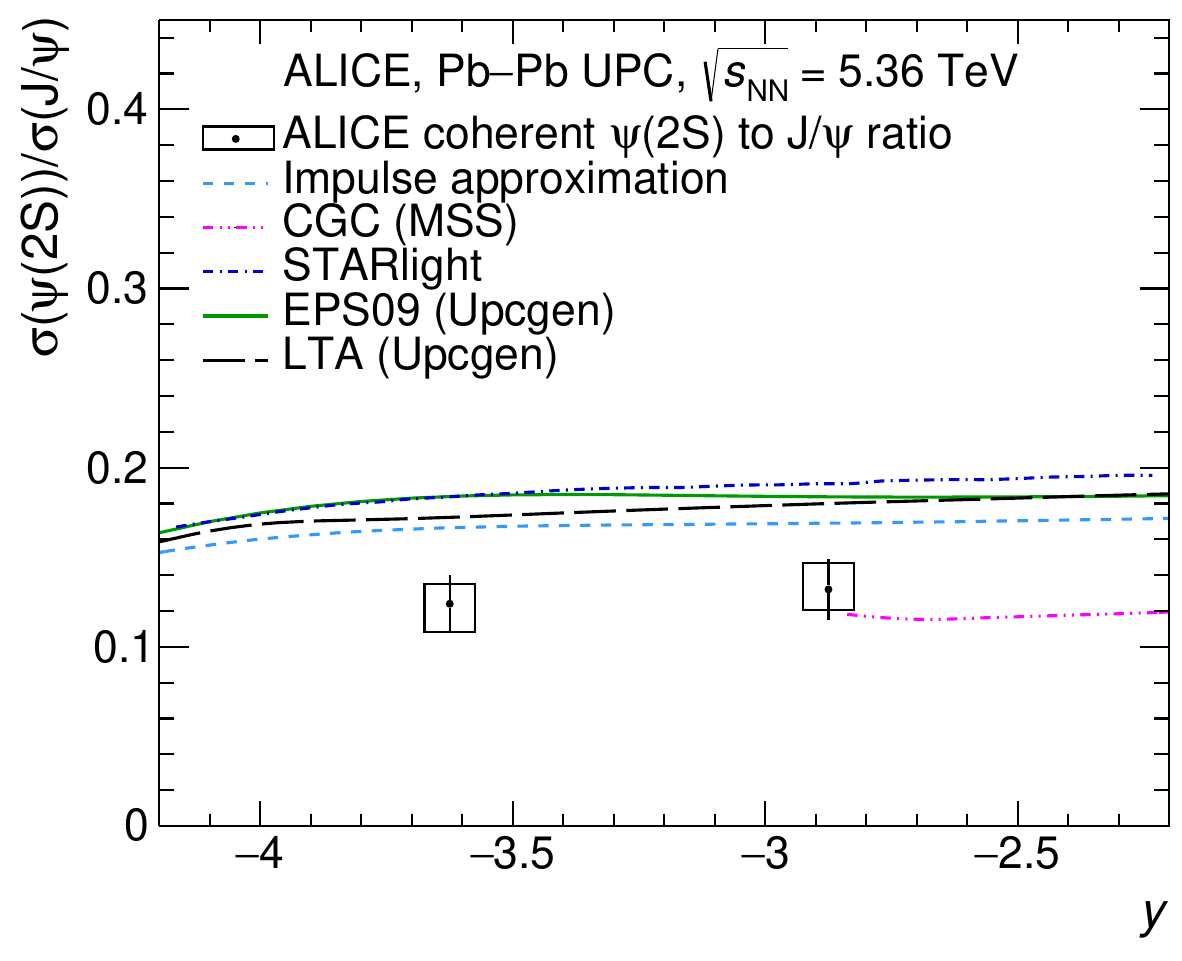}
    \end{minipage}
    \caption{Measured ratio of differential photoproduction cross sections of the coherent \jpsi and \psip in \PbPb UPCs at \fivethreesixnn. The results are compared with theoretical calculations described in the text. The statistical uncertainties are shown as bars, the systematic uncertainties are shown as boxes.}
    \label{fig:cs_vm_ratio}
\end{figure}

Figure~\ref{fig:cs_vm_ratio} compares the measured ratio with theoretical predictions. The CGC-based model is consistent with the measurement within its validity range. The impulse approximation, which uses the \psip-to-\jpsi cross section ratio 
$R = 0.166 \pm 0.007 {\rm (stat.)} \pm 0.008 {\rm (syst.)} \pm 0.007 {\rm (BR)}$ measured by the H1 collaboration~\cite{H1:2002yab}, lies above the ALICE data but remains compatible within 1.9$\sigma$ in the rapidity interval $-4 < y < -3.25$ and within 1.4$\sigma$ in the interval $-3.25 < y < -2.5$. The STARlight, LTA, and EPS09 models predict nearly identical cross section ratios, differing by less than 0.01 and lying slightly above the impulse approximation prediction. All these predictions remain compatible with the data within the present uncertainties.

\subsection{Cross section of exclusive dimuon production}
\label{subsec:results_gl}

The invariant mass-differential cross section for the exclusive dimuon production is extracted for a given invariant mass $\Delta m$ as follows
\begin{equation}
    \frac{\dd \sigma^{\rm excl}_{\gamma\gamma\to\mu\mu}}{\dd m_{\mu\mu}} = \frac{N_{\gamma\gamma\to\mu\mu}}{(1 + \fIgl) \varepsilon_{\rm rec} \varepsilon_{\rm veto} \lumi \Delta m}~.
\end{equation}
The raw $\gamma\gamma\to\mu\mu$ yield values, reconstruction efficiencies $\varepsilon_{\rm rec}$, \fIgl fractions, and the resulting cross sections with corresponding statistical and systematic uncertainties are summarized in 
\Cref{tab:gl_summary} for four rapidity ranges.
\begin{table}[ht]
\centering
\caption{Raw exclusive dimuon production yields with their statistical uncertainties, reconstruction efficiencies, $\fIgl$ fractions and exclusive cross sections.}
\begin{tabular}{ccccccccccc}
\toprule
Rapidity &
$m_{\mu\mu}$ (\GeVmass) & $N_{\gamma\gamma\to\mu\mu}$ & $\varepsilon_{\rm rec}$~(\%) & $\fIgl$ & $\dd\sigma/\dd m$~(mb/\GeVmass) \\ \midrule \\[-0.4cm]
\multirow{5}*[-0.23cm]{$(-4.0, -2.5)$}
& $(1.5, 2.0)$  & $166809 \pm 408$ & $4.9$ & $0.024$ & $5.700 \pm 0.014$~(stat.)~$_{-0.308}^{+0.888}$~(syst.) \\[0.1cm]
& $(2.0, 2.5)$  & $71565 \pm 268$  & $5.4$ & $0.031$ & $2.213 \pm 0.008$~(stat.)~$_{-0.119}^{+0.169}$~(syst.) \\[0.1cm]
& $(2.5, 3.5)$  & $50661 \pm 225$  & $5.4$ & $0.034$ & $0.780 \pm 0.003$~(stat.)~$_{-0.041}^{+0.048}$~(syst.) \\[0.1cm]
& $(3.5, 4.5)$  & $15019 \pm 123$  & $5.1$ & $0.034$ & $0.242 \pm 0.002$~(stat.)~$_{-0.013}^{+0.015}$~(syst.) \\[0.1cm]
& $(4.5, 10.0)$ & $9329 \pm 97$    & $4.4$ & $0.037$ & $0.0321 \pm 0.0003$~(stat.)~$_{-0.0019}^{+0.0023}$~(syst.) \\[0.1cm]
\midrule
\multirow{5}*[-0.23cm]{$(-4.0, -3.5)$}
& $(1.5, 2.0)$  & $35492 \pm 188$ & $4.6$ & $0.033$ & $1.286 \pm 0.007$~(stat.)~$_{-0.071}^{+0.076}$~(syst.) \\[0.1cm]
& $(2.0, 2.5)$  & $11848 \pm 109$ & $4.2$ & $0.031$ & $0.473 \pm 0.004$~(stat.)~$_{-0.025}^{+0.030}$~(syst.) \\[0.1cm]
& $(2.5, 3.5)$  & $7197 \pm 85$   & $3.9$ & $0.047$ & $0.152 \pm 0.002$~(stat.)~$_{-0.009}^{+0.009}$~(syst.) \\[0.1cm]
& $(3.5, 4.5)$  & $1804 \pm 42$   & $3.6$ & $0.047$ & $0.041 \pm 0.001$~(stat.)~$_{-0.002}^{+0.002}$~(syst.) \\[0.1cm]
& $(4.5, 10.0)$ & $837 \pm 29$    & $3.5$ & $0.064$ & $0.0035 \pm 0.0001$~(stat.)~$_{-0.0002}^{+0.0002}$~(syst.) \\[0.1cm]
\midrule
\multirow{5}*[-0.23cm]{$(-3.5, -3.0)$}
& $(1.5, 2.0)$  & $97296 \pm 312$ & $8.2$ & $0.019$ & $1.984 \pm 0.006$~(stat.)~$_{-0.107}^{+0.340}$~(syst.) \\[0.1cm]
& $(2.0, 2.5)$  & $42225 \pm 205$ & $9.2$ & $0.024$ & $0.769 \pm 0.004$~(stat.)~$_{-0.041}^{+0.057}$~(syst.) \\[0.1cm]
& $(2.5, 3.5)$  & $30105 \pm 174$ & $9.3$ & $0.027$ & $0.270 \pm 0.002$~(stat.)~$_{-0.014}^{+0.016}$~(syst.) \\[0.1cm]
& $(3.5, 4.5)$  & $8890 \pm 94$   & $9.0$ & $0.027$ & $0.083 \pm 0.001$~(stat.)~$_{-0.004}^{+0.005}$~(syst.) \\[0.1cm]
& $(4.5, 10.0)$ & $5351 \pm 73$   & $7.9$ & $0.030$ & $0.0102 \pm 0.0001$~(stat.)~$_{-0.0006}^{+0.0007}$~(syst.) \\[0.1cm]
\midrule
\multirow{5}*[-0.23cm]{$(-3.0, -2.5)$}
& $(1.5, 2.0)$  & $34021 \pm 184$ & $2.5$ & $0.031$ & $2.305  \pm  0.012$~(stat.)~$_{-0.126}^{+0.517}$~(syst.) \\[0.1cm]
& $(2.0, 2.5)$  & $17492 \pm 132$ & $3.1$ & $0.046$ & $0.919  \pm  0.007$~(stat.)~$_{-0.051}^{+0.087}$~(syst.) \\[0.1cm]
& $(2.5, 3.5)$  & $13519 \pm 116$ & $3.2$ & $0.036$ & $0.344  \pm  0.003$~(stat.)~$_{-0.018}^{+0.025}$~(syst.) \\[0.1cm]
& $(3.5, 4.5)$  & $4379 \pm 66$   & $3.1$ & $0.036$ & $0.115  \pm  0.002$~(stat.)~$_{-0.006}^{+0.009}$~(syst.) \\[0.1cm]
& $(4.5, 10.0)$ & $3141 \pm 56$   & $2.6$ & $0.026$ & $0.0182 \pm 0.0003$~(stat.)~$_{-0.0011}^{+0.0015}$~(syst.) \\[0.1cm]
\bottomrule
\end{tabular}
\label{tab:gl_summary}
\end{table}

\begin{figure}[ht]
    \centering
    \begin{minipage}{.49\textwidth}
        \centering
        \includegraphics[width=0.99\textwidth]{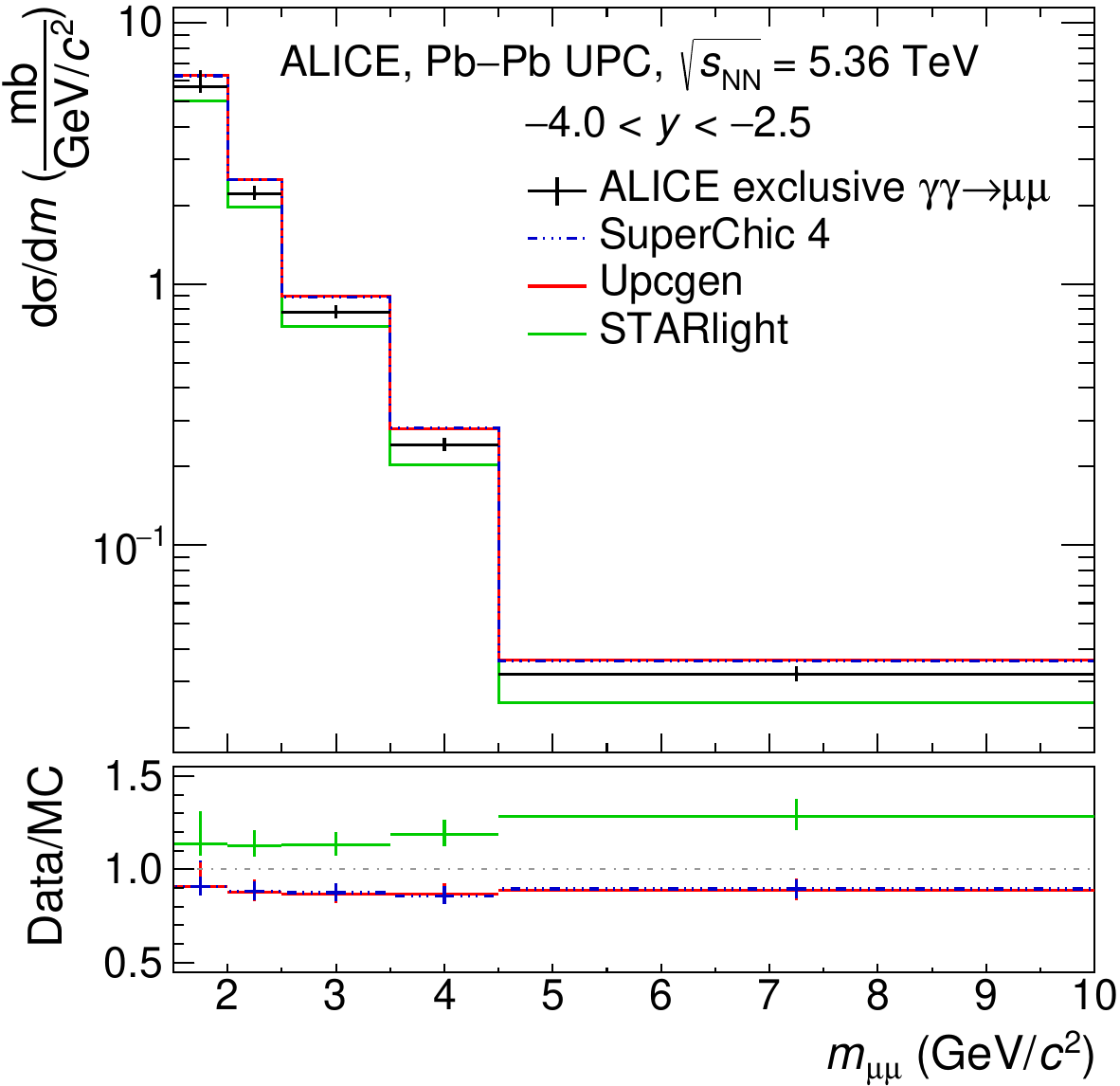}
    \end{minipage}
    \begin{minipage}{.49\textwidth}
        \centering
        \includegraphics[width=0.99\textwidth]{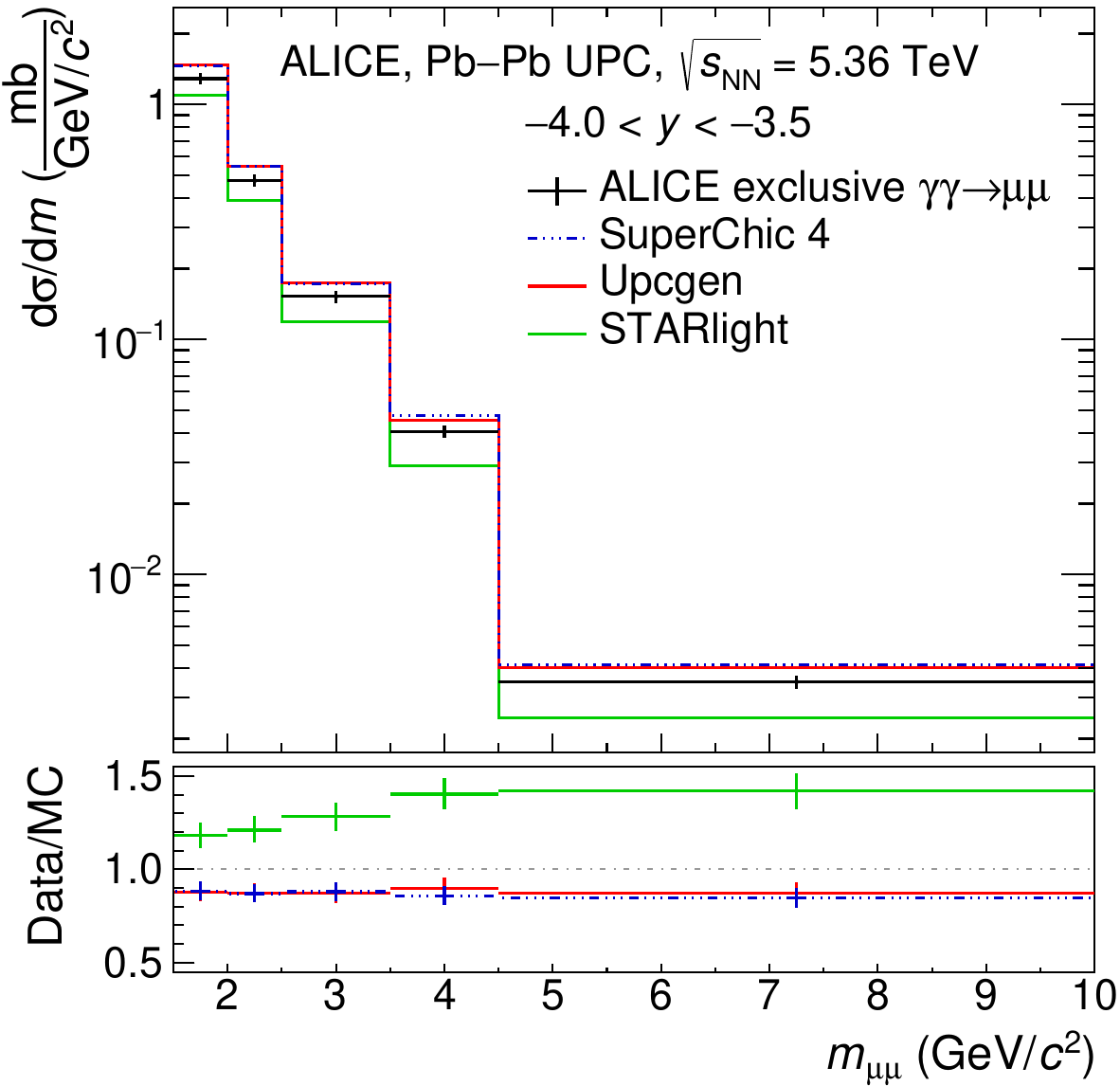}
    \end{minipage}
    \begin{minipage}{.49\textwidth}
        \centering
        \includegraphics[width=0.99\textwidth]{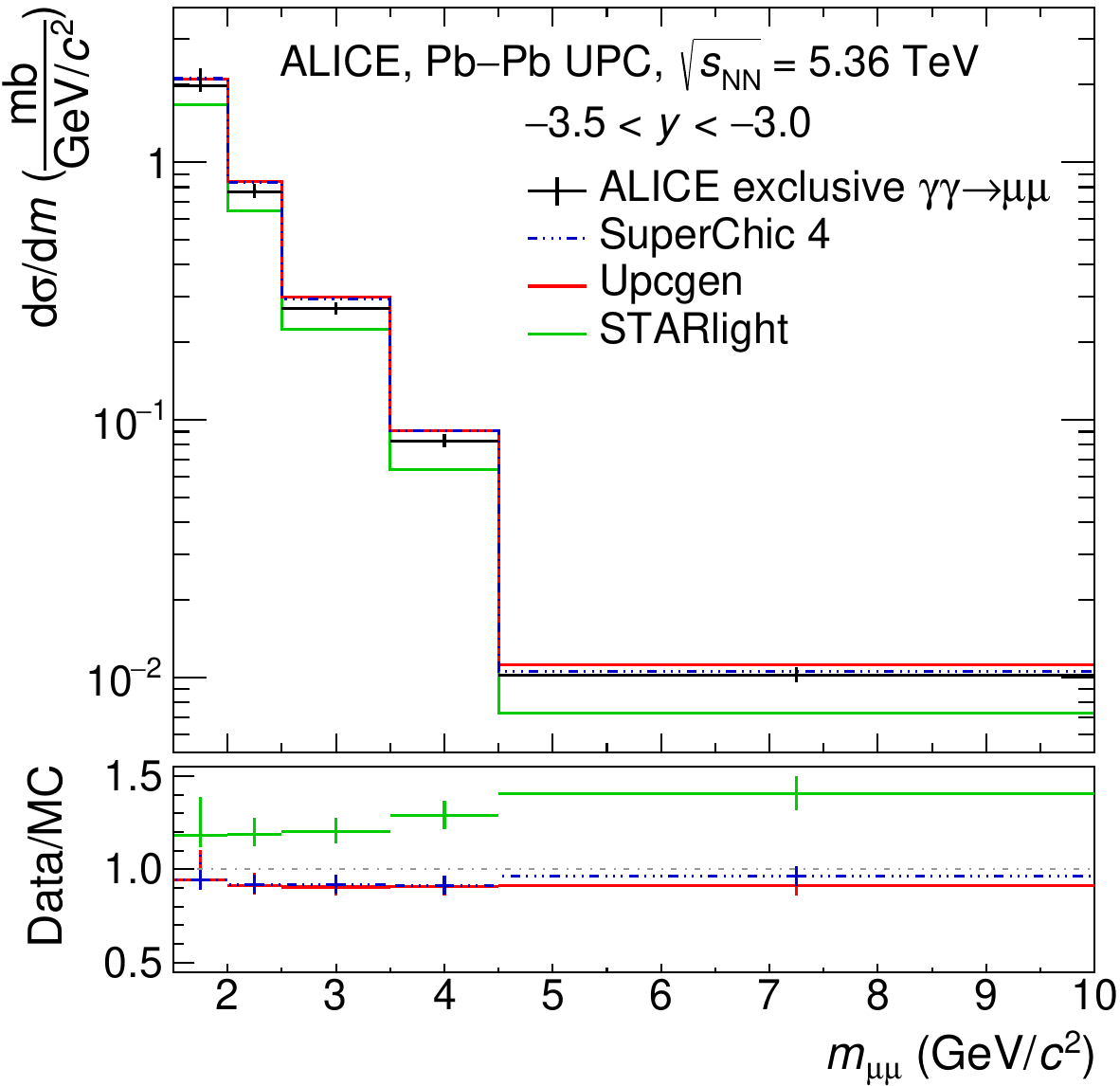}
    \end{minipage}
    \begin{minipage}{.49\textwidth}
        \centering
        \includegraphics[width=0.99\textwidth]{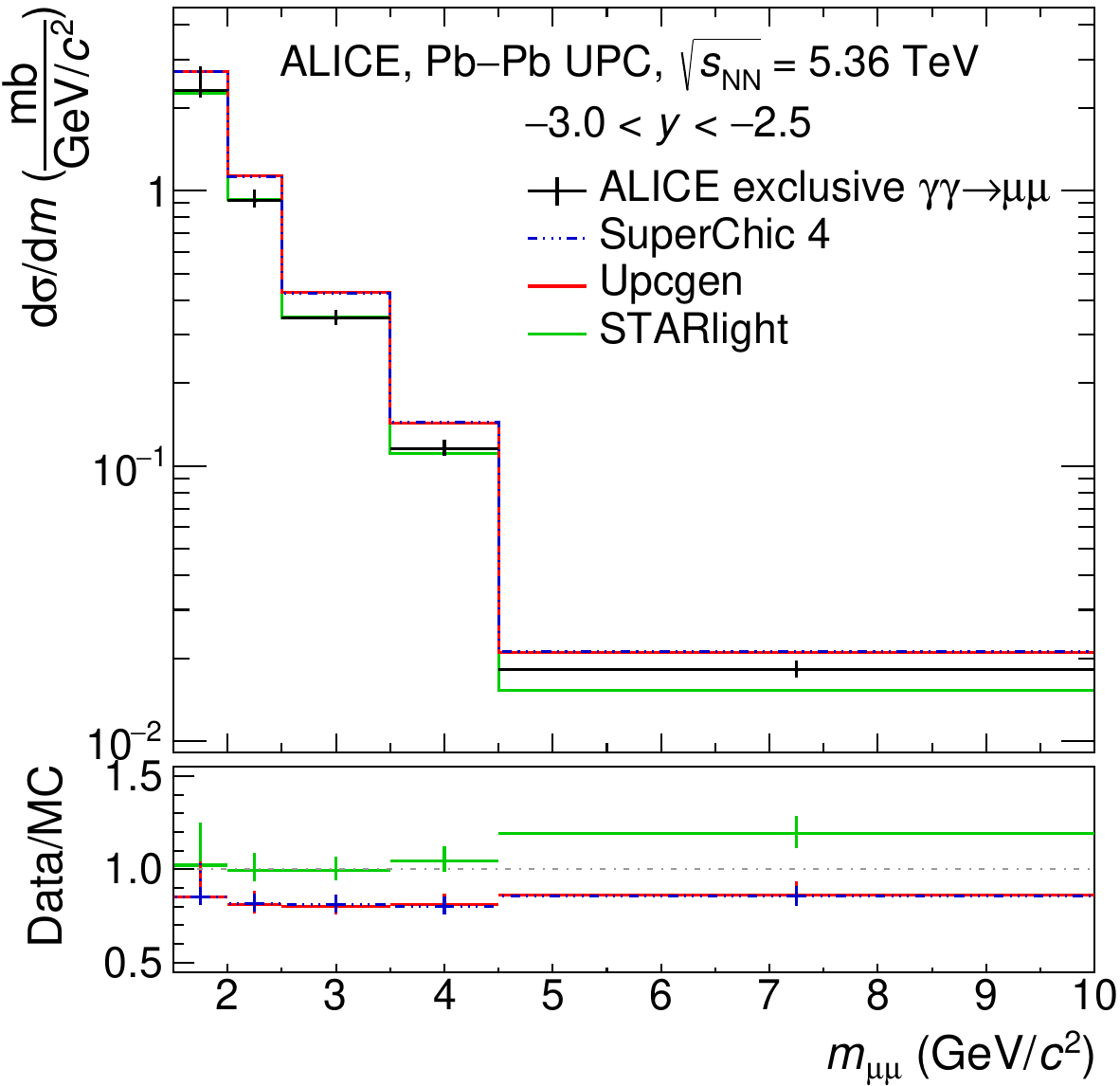}
    \end{minipage}
    \caption{Measured invariant-mass differential cross sections of the exclusive dimuon production $\gamma\gamma\to\mu\mu$ in \PbPb UPCs at \fivethreesixnn. The results are compared with predictions by Monte Carlo generators~\cite{Baltz:2009jk,Klein:2016yzr,Harland-Lang:2018iur,Harland-Lang:2020veo,Burmasov:2021phy,Burmasov:2026ytn} described in the text. The top part of the figures: the absolute values of the cross sections are presented. The bottom part of the figures: the ratios of data to Monte Carlo-based predictions are presented. The bars represent the total statistical and systematic uncertainty added in quadrature.}
    \label{fig:cs_gl}
\end{figure}

Figure~\ref{fig:cs_gl} compares the measured exclusive dimuon production cross sections to predictions from the Upcgen~\cite{Burmasov:2021phy,Burmasov:2026ytn}, SuperChic 4~\cite{Harland-Lang:2018iur,Harland-Lang:2020veo}, and STARlight~\cite{Baltz:2009jk,Klein:2016yzr} generators. STARlight calculates photon fluxes assuming point-like sources with a hard cutoff at the nuclear radius, while Upcgen and SuperChic 4 model photon fluxes using nuclear form factors, which account for photon emission from within the nucleus. This difference leads to higher predicted cross sections in Upcgen and SuperChic 4 compared to STARlight, particularly at high invariant masses and large rapidities, where photon emission at small impact parameters becomes more important. 

STARlight predictions agree with the measured cross sections in the lowest rapidity range $-3.0 < y < -2.5$. However, at higher rapidities, STARlight underestimates the measured exclusive dimuon production cross section by as much as 40\%, the deviation being more pronounced towards higher masses and rapidities. A similar trend is observed in the dimuon and dielectron production measurements at forward rapidity performed by ATLAS~\cite{ATLAS:2020epq, ATLAS:2022srr}. In contrast to STARlight, both Upcgen and SuperChic 4 predictions are approximately 1--2$\sigma$ higher than the data, across all invariant mass intervals and rapidity ranges.

The comparison of the measured cross sections with model predictions highlights the importance of accurate modeling of photon fluxes within the equivalent photon approximation, particularly at high rapidities and invariant masses, where point-like and realistic flux descriptions diverge most significantly~\cite{Burmasov:2021phy}. Furthermore, higher-order QED corrections may be significant, as indicated by theoretical studies~\cite{Zha:2021jhf,Baur:1998ay}, and further high-precision measurements are crucial for providing additional insights.

\section{Conclusion}
\label{sec:conclusion}

Rapidity-differential measurements of the cross sections for the coherent \jpsi and \psip vector meson photoproduction at forward rapidity in \PbPb UPCs at \fivethreesixnn were presented and compared with theoretical models. The impulse approximation significantly overestimates the data by a factor of 1.5--1.8, depending on the rapidity region, confirming the importance of gluon shadowing effects in the nucleus. Leading order pQCD calculations based on the EPS09 LO parametrization or the leading twist approximation give a reasonable description of the measured \psip cross sections, while underestimating the \jpsi data by 1--2$\sigma$. The CGC model agrees within its validity range $|y| < 2.85$, with both the measured \jpsi and \psip cross sections and the corresponding ratios.

Invariant-mass differential cross sections for the exclusive dimuon production at forward rapidity in \PbPb UPCs at \fivethreesixnn were presented and compared with predictions by Monte Carlo event generators in three rapidity regions. The STARlight model, which uses the point-like photon flux approximation, describes the data reasonably well in the $-3 < y < -2.5$ rapidity region, but underestimates the measurements at larger rapidities. Upcgen and SuperChic 4, on the other hand, yield consistently higher cross sections than the measured central values by about 1--2$\sigma$, across all invariant mass and rapidity intervals. The comparison with STARlight, Upcgen, and SuperChic 4 highlights the importance of accurate modeling of photon fluxes within the EPA framework.

Future measurements of coherent heavy vector meson photoproduction with neutron emission at \break\fivethreesixnn will enable more detailed comparisons with theoretical predictions. Furthermore, additional high-precision data for the exclusive dilepton production are needed to further test available models and to investigate the potential relevance of higher-order effects.


\newenvironment{acknowledgement}{\relax}{\relax}
\begin{acknowledgement}
\section*{Acknowledgements}

The ALICE Collaboration would like to thank all its engineers and technicians for their invaluable contributions to the construction of the experiment and the CERN accelerator teams for the outstanding performance of the LHC complex.
The ALICE Collaboration gratefully acknowledges the resources and support provided by all Grid centres and the Worldwide LHC Computing Grid (WLCG) collaboration.
The ALICE Collaboration acknowledges the following funding agencies for their support in building and running the ALICE detector:
A. I. Alikhanyan National Science Laboratory (Yerevan Physics Institute) Foundation (ANSL), State Committee of Science and World Federation of Scientists (WFS), Armenia;
Austrian Academy of Sciences, Austrian Science Fund (FWF): [M 2467-N36] and Nationalstiftung f\"{u}r Forschung, Technologie und Entwicklung, Austria;
Ministry of Communications and High Technologies, National Nuclear Research Center, Azerbaijan;
Rede Nacional de Física de Altas Energias (Renafae), Financiadora de Estudos e Projetos (Finep), Funda\c{c}\~{a}o de Amparo \`{a} Pesquisa do Estado de S\~{a}o Paulo (FAPESP) and The Sao Paulo Research Foundation  (FAPESP), Brazil;
Bulgarian Ministry of Education and Science, within the National Roadmap for Research Infrastructures 2020-2027 (object CERN), Bulgaria;
Ministry of Education of China (MOEC) , Ministry of Science \& Technology of China (MSTC) and National Natural Science Foundation of China (NSFC), China;
Ministry of Science and Education and Croatian Science Foundation, Croatia;
Centro de Aplicaciones Tecnol\'{o}gicas y Desarrollo Nuclear (CEADEN), Cubaenerg\'{\i}a, Cuba;
Ministry of Education, Youth and Sports of the Czech Republic, Czech Republic;
The Danish Council for Independent Research | Natural Sciences, the VILLUM FONDEN and Danish National Research Foundation (DNRF), Denmark;
Helsinki Institute of Physics (HIP), Finland;
Commissariat \`{a} l'Energie Atomique (CEA) and Institut National de Physique Nucl\'{e}aire et de Physique des Particules (IN2P3) and Centre National de la Recherche Scientifique (CNRS), France;
Bundesministerium f\"{u}r Forschung, Technologie und Raumfahrt (BMFTR) and GSI Helmholtzzentrum f\"{u}r Schwerionenforschung GmbH, Germany;
National Research, Development and Innovation Office, Hungary;
Department of Atomic Energy Government of India (DAE), Department of Science and Technology, Government of India (DST), University Grants Commission, Government of India (UGC) and Council of Scientific and Industrial Research (CSIR), India;
National Research and Innovation Agency - BRIN, Indonesia;
Istituto Nazionale di Fisica Nucleare (INFN), Italy;
Japanese Ministry of Education, Culture, Sports, Science and Technology (MEXT) and Japan Society for the Promotion of Science (JSPS) KAKENHI, Japan;
Consejo Nacional de Ciencia (CONACYT) y Tecnolog\'{i}a, through Fondo de Cooperaci\'{o}n Internacional en Ciencia y Tecnolog\'{i}a (FONCICYT) and Direcci\'{o}n General de Asuntos del Personal Academico (DGAPA), Mexico;
Nederlandse Organisatie voor Wetenschappelijk Onderzoek (NWO), Netherlands;
The Research Council of Norway, Norway;
Pontificia Universidad Cat\'{o}lica del Per\'{u}, Peru;
Ministry of Science and Higher Education, National Science Centre and WUT ID-UB, Poland;
Korea Institute of Science and Technology Information and National Research Foundation of Korea (NRF), Republic of Korea;
Ministry of Education and Scientific Research, Institute of Atomic Physics, Ministry of Research and Innovation and Institute of Atomic Physics and Universitatea Nationala de Stiinta si Tehnologie Politehnica Bucuresti, Romania;
Ministerstvo skolstva, vyskumu, vyvoja a mladeze SR, Slovakia;
National Research Foundation of South Africa, South Africa;
Swedish Research Council (VR) and Knut \& Alice Wallenberg Foundation (KAW), Sweden;
European Organization for Nuclear Research, Switzerland;
Suranaree University of Technology (SUT), National Science and Technology Development Agency (NSTDA) and National Science, Research and Innovation Fund (NSRF via PMU-B B05F650021), Thailand;
Turkish Energy, Nuclear and Mineral Research Agency (TENMAK), Turkey;
National Academy of  Sciences of Ukraine, Ukraine;
Science and Technology Facilities Council (STFC), United Kingdom;
National Science Foundation of the United States of America (NSF) and United States Department of Energy, Office of Nuclear Physics (DOE NP), United States of America.
In addition, individual groups or members have received support from:
FORTE project, reg.\ no.\ CZ.02.01.01/00/22\_008/0004632, Czech Republic, co-funded by the European Union, Czech Republic;
European Research Council (grant no. 950692), European Union;
Deutsche Forschungs Gemeinschaft (DFG, German Research Foundation) ``Neutrinos and Dark Matter in Astro- and Particle Physics'' (grant no. SFB 1258), Germany;
CONVECS project, CUP C97H23001700002 FESR 2021-2027 program, Italy.

\end{acknowledgement}

\bibliographystyle{utphys}   
\bibliography{bibliography}

\newpage
\appendix

%
%

\section{The ALICE Collaboration}
\label{app:collab}
\begin{flushleft} 
\small

D.A.H.~Abdallah\,\orcidlink{0000-0003-4768-2718}\,$^{\rm 134}$, 
I.J.~Abualrob\,\orcidlink{0009-0005-3519-5631}\,$^{\rm 112}$, 
S.~Acharya\,\orcidlink{0000-0002-9213-5329}\,$^{\rm 49}$, 
K.~Agarwal\,\orcidlink{0000-0001-5781-3393}\,$^{\rm II,}$$^{\rm 23}$, 
G.~Aglieri Rinella\,\orcidlink{0000-0002-9611-3696}\,$^{\rm 32}$, 
L.~Aglietta\,\orcidlink{0009-0003-0763-6802}\,$^{\rm 24}$, 
N.~Agrawal\,\orcidlink{0000-0003-0348-9836}\,$^{\rm 25}$, 
Z.~Ahammed\,\orcidlink{0000-0001-5241-7412}\,$^{\rm 132}$, 
S.~Ahmad\,\orcidlink{0000-0003-0497-5705}\,$^{\rm 15}$, 
I.~Ahuja\,\orcidlink{0000-0002-4417-1392}\,$^{\rm 36}$
% ...

\section*{Affiliation Notes}

$^{\rm I}$ Deceased\\
$^{\rm II}$ Also at: INFN Trieste\\
$^{\rm III}$ Also at: Fondazione Bruno Kessler (FBK), Trento, Italy\\
$^{\rm IV}$ Also at: Czech Technical University in Prague (CZ)\\
$^{\rm V}$ Also at: Instituto de Fisica da Universidade de Sao Paulo\\
$^{\rm VI}$ Also at: Dipartimento DET del Politecnico di Torino, Turin, Italy\\
$^{\rm VII}$ Also at: Department of Applied Physics, Aligarh Muslim University, Aligarh, India\\
$^{\rm VIII}$ Also at: Institute of Theoretical Physics, University of Wroclaw, Poland\\
$^{\rm IX}$ Also at: Facultad de Ciencias, Universidad Nacional Aut\'{o}noma de M\'{e}xico, Mexico City, Mexico\\

\section*{Collaboration Institutes}

$^{1}$ A.I. Alikhanyan National Science Laboratory (Yerevan Physics Institute) Foundation, Yerevan, Armenia\\
$^{2}$ AGH University of Krakow, Cracow, Poland\\
$^{3}$ Bogolyubov Institute for Theoretical Physics, National Academy of Sciences of Ukraine, Kyiv, Ukraine\\
$^{4}$ Bose Institute, Department of Physics  and Centre for Astroparticle Physics and Space Science (CAPSS), Kolkata, India\\
$^{5}$ California Polytechnic State University, San Luis Obispo, California, United States\\
$^{6}$ Central China Normal University, Wuhan, China\\
$^{7}$ Centro de Aplicaciones Tecnol\'{o}gicas y Desarrollo Nuclear (CEADEN), Havana, Cuba\\
$^{8}$ Centro de Investigaci\'{o}n y de Estudios Avanzados (CINVESTAV), Mexico City and M\'{e}rida, Mexico\\
$^{9}$ Chicago State University, Chicago, Illinois, United States\\
$^{10}$ China Nuclear Data Center, China Institute of Atomic Energy, Beijing, China\\
$^{11}$ China University of Geosciences, Wuhan, China\\
$^{12}$ Chungbuk National University, Cheongju, Republic of Korea\\
$^{13}$ Comenius University Bratislava, Faculty of Mathematics, Physics and Informatics, Bratislava, Slovak Republic\\
$^{14}$ Creighton University, Omaha, Nebraska, United States\\
$^{15}$ Department of Physics, Aligarh Muslim University, Aligarh, India\\
$^{16}$ Department of Physics, Pusan National University, Pusan, Republic of Korea\\
$^{17}$ Department of Physics, Sejong University, Seoul, Republic of Korea\\
$^{18}$ Department of Physics, University of California, Berkeley, California, United States\\
$^{19}$ Department of Physics, University of Oslo, Oslo, Norway\\
$^{20}$ Department of Physics and Technology, University of Bergen, Bergen, Norway\\
$^{21}$ Dipartimento di Fisica, Universit\`{a} di Pavia, Pavia, Italy\\
$^{22}$ Dipartimento di Fisica dell'Universit\`{a} and Sezione INFN, Cagliari, Italy\\
$^{23}$ Dipartimento di Fisica dell'Universit\`{a} and Sezione INFN, Trieste, Italy\\
$^{24}$ Dipartimento di Fisica dell'Universit\`{a} and Sezione INFN, Turin, Italy\\
$^{25}$ Dipartimento di Fisica e Astronomia dell'Universit\`{a} and Sezione INFN, Bologna, Italy\\
$^{26}$ Dipartimento di Fisica e Astronomia dell'Universit\`{a} and Sezione INFN, Catania, Italy\\
$^{27}$ Dipartimento di Fisica e Astronomia dell'Universit\`{a} and Sezione INFN, Padova, Italy\\
$^{28}$ Dipartimento di Fisica `E.R.~Caianiello' dell'Universit\`{a} and Gruppo Collegato INFN, Salerno, Italy\\
$^{29}$ Dipartimento DISAT del Politecnico and Sezione INFN, Turin, Italy\\
$^{30}$ Dipartimento di Scienze MIFT, Universit\`{a} di Messina, Messina, Italy\\
$^{31}$ Dipartimento Interateneo di Fisica `M.~Merlin' and Sezione INFN, Bari, Italy\\
$^{32}$ European Organization for Nuclear Research (CERN), Geneva, Switzerland\\
$^{33}$ Faculty of Electrical Engineering, Mechanical Engineering and Naval Architecture, University of Split, Split, Croatia\\
$^{34}$ Faculty of Nuclear Sciences and Physical Engineering, Czech Technical University in Prague, Prague, Czech Republic\\
$^{35}$ Faculty of Physics, Sofia University, Sofia, Bulgaria\\
$^{36}$ Faculty of Science, P.J.~\v{S}af\'{a}rik University, Ko\v{s}ice, Slovak Republic\\
$^{37}$ Faculty of Technology, Environmental and Social Sciences, Bergen, Norway\\
$^{38}$ Frankfurt Institute for Advanced Studies, Johann Wolfgang Goethe-Universit\"{a}t Frankfurt, Frankfurt, Germany\\
$^{39}$ Fudan University, Shanghai, China\\
$^{40}$ Gauhati University, Department of Physics, Guwahati, India\\
$^{41}$ Helmholtz-Institut f\"{u}r Strahlen- und Kernphysik, Rheinische Friedrich-Wilhelms-Universit\"{a}t Bonn, Bonn, Germany\\
$^{42}$ Helsinki Institute of Physics (HIP), Helsinki, Finland\\
$^{43}$ High Energy Physics Group,  Universidad Aut\'{o}noma de Puebla, Puebla, Mexico\\
$^{44}$ Horia Hulubei National Institute of Physics and Nuclear Engineering, Bucharest, Romania\\
$^{45}$ HUN-REN Wigner Research Centre for Physics, Budapest, Hungary\\
$^{46}$ Indian Institute of Technology Bombay (IIT), Mumbai, India\\
$^{47}$ Indian Institute of Technology Indore, Indore, India\\
$^{48}$ INFN, Laboratori Nazionali di Frascati, Frascati, Italy\\
$^{49}$ INFN, Sezione di Bari, Bari, Italy\\
$^{50}$ INFN, Sezione di Bologna, Bologna, Italy\\
$^{51}$ INFN, Sezione di Cagliari, Cagliari, Italy\\
$^{52}$ INFN, Sezione di Catania, Catania, Italy\\
$^{53}$ INFN, Sezione di Padova, Padova, Italy\\
$^{54}$ INFN, Sezione di Pavia, Pavia, Italy\\
$^{55}$ INFN, Sezione di Torino, Turin, Italy\\
$^{56}$ INFN, Sezione di Trieste, Trieste, Italy\\
$^{57}$ Inha University, Incheon, Republic of Korea\\
$^{58}$ Institute for Gravitational and Subatomic Physics (GRASP), Utrecht University/Nikhef, Utrecht, Netherlands\\
$^{59}$ Institute of Experimental Physics, Slovak Academy of Sciences, Ko\v{s}ice, Slovak Republic\\
$^{60}$ Institute of Physics, Homi Bhabha National Institute, Bhubaneswar, India\\
$^{61}$ Institute of Physics of the Czech Academy of Sciences, Prague, Czech Republic\\
$^{62}$ Institute of Space Science (ISS), Bucharest, Romania\\
$^{63}$ Institut f\"{u}r Kernphysik, Johann Wolfgang Goethe-Universit\"{a}t Frankfurt, Frankfurt, Germany\\
$^{64}$ Instituto de Ciencias Nucleares, Universidad Nacional Aut\'{o}noma de M\'{e}xico, Mexico City, Mexico\\
$^{65}$ Instituto de F\'{i}sica, Universidade Federal do Rio Grande do Sul (UFRGS), Porto Alegre, Brazil\\
$^{66}$ Instituto de F\'{\i}sica, Universidad Nacional Aut\'{o}noma de M\'{e}xico, Mexico City, Mexico\\
$^{67}$ iThemba LABS, National Research Foundation, Somerset West, South Africa\\
$^{68}$ Jeonbuk National University, Jeonju, Republic of Korea\\
$^{69}$ Korea Institute of Science and Technology Information, Daejeon, Republic of Korea\\
$^{70}$ Laboratoire de Physique Subatomique et de Cosmologie, Universit\'{e} Grenoble-Alpes, CNRS-IN2P3, Grenoble, France\\
$^{71}$ Lawrence Berkeley National Laboratory, Berkeley, California, United States\\
$^{72}$ Lund University Department of Physics, Division of Particle Physics, Lund, Sweden\\
$^{73}$ Marietta Blau Institute, Vienna, Austria\\
$^{74}$ Nagasaki Institute of Applied Science, Nagasaki, Japan\\
$^{75}$ Nara Women{'}s University (NWU), Nara, Japan\\
$^{76}$ National Centre for Nuclear Research, Warsaw, Poland\\
$^{77}$ National Institute of Science Education and Research, Homi Bhabha National Institute, Jatni, India\\
$^{78}$ National Nuclear Research Center, Baku, Azerbaijan\\
$^{79}$ National Research and Innovation Agency - BRIN, Jakarta, Indonesia\\
$^{80}$ Niels Bohr Institute, University of Copenhagen, Copenhagen, Denmark\\
$^{81}$ Nikhef, National institute for subatomic physics, Amsterdam, Netherlands\\
$^{82}$ Nuclear Physics Group, STFC Daresbury Laboratory, Daresbury, United Kingdom\\
$^{83}$ Nuclear Physics Institute of the Czech Academy of Sciences, Husinec-\v{R}e\v{z}, Czech Republic\\
$^{84}$ Oak Ridge National Laboratory, Oak Ridge, Tennessee, United States\\
$^{85}$ Ohio State University, Columbus, Ohio, United States\\
$^{86}$ Physics department, Faculty of science, University of Zagreb, Zagreb, Croatia\\
$^{87}$ Physics Department, Panjab University, Chandigarh, India\\
$^{88}$ Physics Department, University of Jammu, Jammu, India\\
$^{89}$ Physics Program and International Institute for Sustainability with Knotted Chiral Meta Matter (WPI-SKCM$^{2}$), Hiroshima University, Hiroshima, Japan\\
$^{90}$ Physikalisches Institut, Eberhard-Karls-Universit\"{a}t T\"{u}bingen, T\"{u}bingen, Germany\\
$^{91}$ Physikalisches Institut, Ruprecht-Karls-Universit\"{a}t Heidelberg, Heidelberg, Germany\\
$^{92}$ Physik Department, Technische Universit\"{a}t M\"{u}nchen, Munich, Germany\\
$^{93}$ Politecnico di Bari and Sezione INFN, Bari, Italy\\
$^{94}$ Research Division and ExtreMe Matter Institute EMMI, GSI Helmholtzzentrum f\"ur Schwerionenforschung GmbH, Darmstadt, Germany\\
$^{95}$ Saga University, Saga, Japan\\
$^{96}$ Saha Institute of Nuclear Physics, Homi Bhabha National Institute, Kolkata, India\\
$^{97}$ School of Physics and Astronomy, University of Birmingham, Birmingham, United Kingdom\\
$^{98}$ Secci\'{o}n F\'{\i}sica, Departamento de Ciencias, Pontificia Universidad Cat\'{o}lica del Per\'{u}, Lima, Peru\\
$^{99}$ SUBATECH, IMT Atlantique, Nantes Universit\'{e}, CNRS-IN2P3, Nantes, France\\
$^{100}$ Sungkyunkwan University, Suwon City, Republic of Korea\\
$^{101}$ Suranaree University of Technology, Nakhon Ratchasima, Thailand\\
$^{102}$ Technical University of Ko\v{s}ice, Ko\v{s}ice, Slovak Republic\\
$^{103}$ The Henryk Niewodniczanski Institute of Nuclear Physics, Polish Academy of Sciences, Cracow, Poland\\
$^{104}$ The University of Texas at Austin, Austin, Texas, United States\\
$^{105}$ Universidad Aut\'{o}noma de Sinaloa, Culiac\'{a}n, Mexico\\
$^{106}$ Universidade de S\~{a}o Paulo (USP), S\~{a}o Paulo, Brazil\\
$^{107}$ Universidade Estadual de Campinas (UNICAMP), Campinas, Brazil\\
$^{108}$ Universidade Federal do ABC, Santo Andre, Brazil\\
$^{109}$ Universitatea Nationala de Stiinta si Tehnologie Politehnica Bucuresti, Bucharest, Romania\\
$^{110}$ University of Cape Town, Cape Town, South Africa\\
$^{111}$ University of Derby, Derby, United Kingdom\\
$^{112}$ University of Houston, Houston, Texas, United States\\
$^{113}$ University of Jyv\"{a}skyl\"{a}, Jyv\"{a}skyl\"{a}, Finland\\
$^{114}$ University of Kansas, Lawrence, Kansas, United States\\
$^{115}$ University of Liverpool, Liverpool, United Kingdom\\
$^{116}$ University of Science and Technology of China, Hefei, China\\
$^{117}$ University of Silesia in Katowice, Katowice, Poland\\
$^{118}$ University of South-Eastern Norway, Kongsberg, Norway\\
$^{119}$ University of Tennessee, Knoxville, Tennessee, United States\\
$^{120}$ University of the Witwatersrand, Johannesburg, South Africa\\
$^{121}$ University of Tokyo, Tokyo, Japan\\
$^{122}$ University of Tsukuba, Tsukuba, Japan\\
$^{123}$ Universit\"{a}t M\"{u}nster, Institut f\"{u}r Kernphysik, M\"{u}nster, Germany\\
$^{124}$ Universit\'{e} Clermont Auvergne, CNRS/IN2P3, LPC, Clermont-Ferrand, France\\
$^{125}$ Universit\'{e} de Lyon, CNRS/IN2P3, Institut de Physique des 2 Infinis de Lyon, Lyon, France\\
$^{126}$ Universit\'{e} de Strasbourg, CNRS, IPHC UMR 7178, F-67000 Strasbourg, France, Strasbourg, France\\
$^{127}$ Universit\'{e} Paris-Saclay, Centre d'Etudes de Saclay (CEA), IRFU, D\'{e}partment de Physique Nucl\'{e}aire (DPhN), Saclay, France\\
$^{128}$ Universit\'{e}  Paris-Saclay, CNRS/IN2P3, IJCLab, Orsay, France\\
$^{129}$ Universit\`{a} degli Studi di Foggia, Foggia, Italy\\
$^{130}$ Universit\`{a} del Piemonte Orientale, Vercelli, Italy\\
$^{131}$ Universit\`{a} di Brescia, Brescia, Italy\\
$^{132}$ Variable Energy Cyclotron Centre, Homi Bhabha National Institute, Kolkata, India\\
$^{133}$ Warsaw University of Technology, Warsaw, Poland\\
$^{134}$ Wayne State University, Detroit, Michigan, United States\\
$^{135}$ Yale University, New Haven, Connecticut, United States\\
$^{136}$ Yildiz Technical University, Istanbul, Turkey\\
$^{137}$ Yonsei University, Seoul, Republic of Korea\\
$^{138}$ Affiliated with an institute formerly covered by a cooperation agreement with CERN\\
$^{139}$ Affiliated with an international laboratory covered by a cooperation agreement with CERN.\\

\end{flushleft} 

\end{document}